\documentclass{article}
\usepackage[margin=1in,paperheight=11in,paperwidth=8.5in]{geometry}
\usepackage{graphicx}
\usepackage{authblk}
\usepackage{mathtools}
\usepackage{amssymb} 
\usepackage{float}
\usepackage{enumitem}
\usepackage{amsthm}
\usepackage{amsmath}
\usepackage{amsfonts}
\usepackage{array}
\usepackage{amsopn}
\usepackage{multirow}
\usepackage{longtable}
 \usepackage[table,xcdraw]{xcolor}
\usepackage{natbib}
\usepackage{enumitem}
\DeclarePairedDelimiterX\Basics[1](){ #1}
\usepackage[utf8]{inputenc}
\usepackage{color}
\usepackage{caption}
\usepackage{subfigure}
\usepackage{enumitem}
\usepackage{pifont}
\usepackage{algorithm, algpseudocode, algcompatible}
\usepackage[T1]{fontenc}
\usepackage{etoolbox}
\usepackage{arydshln}
\usepackage{lscape}
\usepackage{bm}
\usepackage{multirow}
\usepackage{lineno}
\usepackage{diagbox}
\usepackage[flushleft]{threeparttable}
\usepackage{array, booktabs, multirow}
\newcommand{\cmark}{\ding{51}}%
\newcommand{\xmark}{\ding{55}}%

\usepackage{setspace}
\DeclareCaptionJustification{double}{\doublespacing}
\usepackage[labelfont=bf]{caption}

\newcolumntype{C}[1]{>{\centering\arraybackslash}p{#1}}

\usepackage[colorlinks=true,linkcolor=red,filecolor=green,citecolor=blue]{hyperref}

\usepackage{tcolorbox}
\newcommand*\samethanks[1][\value{footnote}]{\footnotemark[#1]}

\newcommand{\univ}{\textbf{\textsc{Universal}}}
\newcommand{\single}{\textbf{\textsc{Single}}}
\newcommand{\aug}{\textbf{\textsc{Augmented}}}
\definecolor{aoe}{rgb}{0.0, 0.5, 0.0}

\usepackage[normalem]{ulem}

\title{\textbf{Volatility Forecasting with Machine Learning \\ and Intraday Commonality}}

\author[1,2]{Chao Zhang\thanks{We would like to thank two anonymous referees, the associate editor and the editor, Dacheng Xiu, for their valuable comments. We are grateful to Rama Cont, Alvaro Cartea, Blanka Horvath, participants at the 11th Bachelier World Congress 2022, the 2022 Asian Finance Association Annual Conference for helpful comments. The authors also thank the Oxford Suzhou Centre for Advanced Research for providing the computational facility. The first two authors contributed equally to this work. The first author acknowledges the support from Clarendon Fund. The second author acknowledges the support from EPSRC Centre for Doctoral Training in Mathematics of Random Systems: Analysis, Modelling and Simulation (EP/S023925/1). Correspondence to: Chao Zhang <chao.zhang@stats.ox.ac.uk>.}}
\author[2,4]{Yihuang Zhang\samethanks}
\author[1,2,3]{Mihai Cucuringu}
\author[2,4]{Zhongmin Qian}

\affil[1]{\small Department of Statistics, University of Oxford, Oxford, UK}
\affil[2]{\small Mathematical Institute, University of Oxford, Oxford, UK}
\affil[3]{\small The Alan Turing Institute, London, UK}
\affil[4]{\small Oxford Suzhou Centre for Advanced Research}

\begin{document}
\maketitle
\begin{abstract}
We apply machine learning models to forecast intraday realized volatility (RV), by exploiting commonality in intraday volatility via pooling stock data together, and by incorporating a proxy for the market volatility. Neural networks dominate linear regressions and tree-based models in terms of performance, due to their ability to uncover and model complex latent interactions among variables. Our findings remain robust when we apply trained models to new stocks that have not been included in the training set, thus providing new empirical evidence for a universal volatility mechanism among stocks. Finally, we propose a new approach to forecasting one-day-ahead RVs using past intraday RVs as predictors, and highlight interesting time-of-day effects that aid the forecasting mechanism. The results demonstrate that the proposed methodology yields superior out-of-sample forecasts over a strong set of traditional baselines that only rely on past daily RVs.
\end{abstract}

\hfill


\newpage

\section{Introduction} \label{sec:intro}

Forecasting and modeling stock return volatility has been of interest to both academics and practitioners. Recent advances in high-frequency trading (HFT)  highlight the need for robust and accurate intraday volatility forecasts. {For example, Deutsche B\"{o}rse, one of the world's leading data and technology service providers, launched the ``Intraday Volatility Forecast'' project in 2015 to provide intraday volatility forecasts up to 30-min for DAX, EURO STOXX 50 and Euro-Bund.}

{\citet{engle2012forecasting} pointed out that intraday volatility forecasts are important for managing risk, pricing derivatives, and devising quantitative strategies, especially in high-frequency trading. \citet{stroud2014bayesian} also demonstrated that intraday measures are useful for market makers, high-frequency trading, and option traders. Specifically, intraday volatility forecasts may support traders in assessing the likelihood of price changes and therefore better understanding the risk involved in certain automated trading strategies (see \citet{bates2019crashes}). Screen traders could leverage volatility forecasts to support live trading and enhance the pre-trade transaction cost analysis from the risk assessment of price slippage. Intraday volatility forecasts are also helpful for practitioners screening for high-volatility opportunities and trading corresponding option strategies (see \citet{ni2008volatility}).}

{To the best of our knowledge}, unlike daily volatility forecasting, intraday volatility has not yet received much attention in the research literature. It is pointed out by \citet{andersen1997intraday} that conventional parametric models, such as GARCH and stochastic volatility (SV) models, may fail to reveal certain  features of intraday returns. In \citet{andersen2003modeling} and \citet{corsi2009simple}, high-frequency data are used to estimate daily realized volatility (RV) by summing squared intraday returns. Some related literature on this aspect will be reviewed briefly in the next section.  These methods {may reveal important information about characteristics of daily returns and volatilities, but do not easily lend themselves applicable to the task of}   forecasting intraday volatility. 

{In the present paper, we study several non-parametric machine learning (ML) models for forecasting \textit{multi-asset intraday volatility} by leveraging high-frequency data from the U.S. equity market. We first propose a measure for evaluating the commonality in intraday volatility. The results demonstrate that, by taking advantage of commonality in intraday volatility, the forecasting performance of these ML models improves significantly.
Neural networks yield both statistically and economically significant improvements in out-of-sample performance over linear regressions and tree-based models, due to their ability to uncover the non-linearity and model complex latent interactions among variables.
The improvements remain robust when we apply trained models to new stocks that have not been included in the training set, thus alleviating the overfitting concerns of NNs and providing new evidence towards a certain universality phenomenon in modeling volatility. In the end, our findings reveal that past intraday volatilities provide additional useful information for forecasting daily volatility, and reveal subtle time-of-day effects that aid the forecasting mechanism.}

We augment our proposed methodology with a very thorough set of numerical experiments. The data covered in this work spans the period from July 2011 to June 2021, and includes the top 100 {most liquid} components of the S\&P 500 index, and the 10-min, 30-min, 65-min, and daily (without overnight information) forecasting horizons are analyzed.

More specifically, a measure for evaluating the commonality in intraday volatility is proposed, that is the adjusted R-squared value from linear regressions of the RVs of a given stock against the market RVs. It is demonstrated that commonality over the daily horizon is turbulent over time, although the commonality in intraday RVs is strong and stable. The analysis of the high-frequency data from the real market reveals the following interesting phenomena.
During a trading session, commonality achieves a peak near closing sessions, in contrast to the diurnal volatility pattern.

Second, in order to assess the benefits of incorporating commonality into models used to predict intraday volatility, multiple machine learning algorithms (including ARIMA, HAR, OLS, LASSO, XGBoost, MLP, and LSTM) are implemented under three different schemes: (a) {\single}: training specific models for each asset; (b) {\univ}: training one model with pooled data for all assets; (c) {\aug}:  training one model using pooled data with an additional predictor which takes into  account the impact of market realized volatility. It is revealed that for most models, the incorporation of intraday commonality likely leads to better out-of-sample performance, based on the pooled data together with additional information of the market volatility.

The empirical results we present in the paper demonstrate that neural networks (NNs) can be {superior to other techniques}. Empirical evidence is provided to demonstrate the capability of NNs for capturing complex interactions among predictors. Furthermore, to alleviate the concerns of over-fitting, a stringent out-of-sample test is conducted, where the  trained models are evaluated on completely new stocks which have  not been included in the training sample. {Our results reveal that}  NNs can outperform other approaches. By comparing the result with the performance obtained by OLS models trained for each new stock, we show the validity of a universal volatility mechanism among stocks. 
Similar findings are reported in \citet{sirignano2019universal} {concerning universal features of price formation in equity markets}.

We conclude the paper by proposing a new approach for predicting daily volatility, in which the past intraday volatilities rather than the past daily volatilities are used as predictors. This approach fully utilizes the available high-frequency data, and therefore contributes to the improvement over traditional methods of modeling  daily volatilities. The results presented in this paper demonstrate that machine learning models, where past intraday  volatilities are used as predictors, {tend} to outperform the traditional models with past daily volatilities (e.g. HAR of \citet{corsi2009simple}, SHAR of \citet{patton2015good}, HARQ of \citet{bollerslev2016exploiting}). 
{We believe that our approach brings a novel perspective on research that studies the effectiveness of past intraday volatilities in forecasting future daily volatility, providing new insights into the understanding of volatility dynamics.}  

The remainder of this paper is structured as follows. We begin with Section \ref{sec:literature} by reviewing some closely related literature, which however should not be considered as a comprehensive survey of the subject. 
In Section \ref{sec:data}, the data and the definition of realized volatility are described. In Section \ref{sec:commonality}, we discuss the commonality in intraday volatility. Various machine learning models and three training schemes for predicting future intraday volatility are introduced in Section \ref{sec:methodology}. Section \ref{sec:exp} provides the forecasting results and discusses the empirical findings. In Section \ref{sec:intra2daily}, a new approach to forecasting daily volatility using past intraday volatility as predictors is proposed. Finally, we summarize our study and discuss further avenues of investigation in Section \ref{sec:conclusion}.

\section{Related literature}\label{sec:literature}

Our study is built upon several research streams proposed by various authors over the recent years. 
The first stream is related to the research on the commonality in financial markets. \citet{chordia2000commonality} have recognized the existence of commonality in liquidity, and \citet{karolyi2012understanding} have suggested that commonality in liquidity is related to market volatility, in particular, the presence of international investors and trading activity. \citet{dang2015commonality} have made an observation that the news commonality is associated with stock return co-movement and liquidity commonality.

The co-movement in daily volatility is well-known from the previous literature. Traditional GARCH and stochastic volatility models (e.g. \citet{andersen2006volatility,calvet2006volatility}) all make use of the volatility spillover effects. \citet{herskovic2016common} have provided empirical evidence of the co-movement in volatility across the equity market. \citet{bollerslev2018risk} have observed strong similarities in daily realized volatility and have utilized them to forecast the daily realized volatility. \citet{engle2012forecasting} have emphasized that pooled data is useful for intraday volatility forecasting, and \citet{herskovic2020firm} have reported that volatilities co-move strongly over time. However, there is still a void of research related to commonality in intraday volatility and its implications for managing intraday risks, especially for forecasting purposes.

Second, there are numerous contributions in the existing literature on the topic of forecasting daily volatility. However, most methods proposed by various researchers for modeling and forecasting return volatility largely rely on the parametric GARCH or stochastic volatility models, which provide forecasts of daily volatility from daily return. As pointed out by  \citet{andersen2003modeling, andersen2006volatility, engle2007good}, these models employed to predict daily volatility cannot take advantage of high-frequency data, and suffer from the curse of high-dimensionality when dealing with multiple assets simultaneously. Due to the availability of high-frequency data, realized  volatility (RV), computed from summing squared intraday returns, has gained popularity in recent years. \citet{andersen2003modeling} have proposed an ARFIMA model for forecasting daily RVs, which outperforms conventional GARCH and related approaches. 
\citet{corsi2009simple} has put forward a parsimonious AR-type model, termed Heterogeneous Autoregressive (HAR), for predicting daily RVs using various realized volatility components over different time horizons. 
Recently \citet{izzeldin2019forecasting} have made a comparison investigation for the forecasting performance of ARFIMA and HAR, and have concluded their performance {is essentially indistinguishable}. See Section \ref{sec:intra2daily} for more models to predict daily volatility.

Nonetheless, little attention has been paid to the role of forecasting intraday volatility.
\citet{taylor1997incremental} has proposed an hourly volatility model based on an ARCH specification, and  \citet{engle2012forecasting} have constructed a GARCH model for intraday financial returns, by specifying the variance as a product of daily, diurnal, and stochastic intraday components. {These models, such as  traditional GARCH and SV, are potentially restrictive due to their parametric nature, 
and are not able to effectively take into account the non-linear and highly complex relationships among different financial variables.}

Third, machine learning models have demonstrated great potential in finance, such as their applications in asset pricing. The high-dimensional nature of ML methods allows for better approximations to unknown and potentially complex data-generating processes, in contrast with traditional economic models.  \citet{gu2020empirical} have pointed out the superior performance of ML models for empirical asset pricing. Recently, \citet{xiong2015deep} have applied LSTMs to forecast S\&P 500 volatility, with Google domestic trends as predictors, and \citet{bucci2020realized} has demonstrated that RNNs are able to outperform all the traditional econometric methods in forecasting monthly volatility of the S\&P index. \citet{rahimikia2020machine} have compared machine learning models with HAR models for forecasting daily realized volatility by using variables extracted from limit order books and news. 
\citet{li2020forecasting} have proposed a simple average ensemble model 
combining multiple machine learning algorithms for forecasting daily (and  monthly) realized volatility, and \citet{christensen2021machine} have examined the performance of machine learning models in forecasting one-day-ahead realized volatility with firm-specific characteristics and macroeconomic indicators.

\section{Data and realized volatility} \label{sec:data}
\subsection{Data} 
We use the Nasdaq ITCH data from LOBSTER\footnote{https://lobsterdata.com/} to compute intraday returns via mid-prices. 
We select the top 100 components of S\&P 500 index, for the period between 2011-07-01 and 2021-06-30. After filtering out the stocks for which the dataset does not span the entire sample period, we are left with 93 stocks.
Table \ref{tab:sector} presents the number of stocks in each sector, according to the GICS sector division.\footnote{The Global Industry Classification Standard (GICS) is an industry taxonomy developed in 1999 by MSCI and Standard \& Poor's (S\&P).} 

\begin{table}[H]
\centering
\caption{Components in each sector.}
\resizebox{1.0\textwidth}{!}{\begin{tabular}{ccc}
   \toprule
Sector                 & Number & Tickers  \\
   \midrule
\multirow{2}{*}{Information Technology} & \multirow{2}{*}{20}     & AAPL ACN ADBE ADP AVGO CRM CSCO FIS FISV IBM \\
& & INTC INTU MA MSFT MU NVDA   ORCL QCOM TXN V \\
\multirow{2}{*}{Health Care}            &  \multirow{2}{*}{19}     & ABT AMGN BDX BMY BSX CI CVS DHR GILD ISRG \\
& & JNJ LLY MDT MRK PFE SYK TMO UNH   VRTX         \\
\multirow{2}{*}{Financials}             &  \multirow{2}{*}{15}     & AXP BAC BLK BRK.B C CB CME GS JPM MMC \\
&& MS PNC SCHW USB WFC      \\  
Industrials            & 9      & BA CAT CSX GE HON LMT MMM UNP UPS \\
Consumer Discretionary & 8      & AMZN HD LOW MCD NKE SBUX TGT TJX \\
Consumer Staples       & 8      & CL COST KO MO PEP PG PM WMT   \\
Communication Services & 6      & CMCSA DIS GOOG NFLX T VZ    \\
Others                 & 8      & AMT CCI COP CVX D DUK SO XOM   \\
\bottomrule
\end{tabular}}
\label{tab:sector}
\end{table}

\subsection{Realized volatility}
In a general form,  $P_{i,t}$ denotes the price process of a financial asset $i$ and it follows
\begin{equation}\label{eq:price}
\mathrm{d} \operatorname{log} P_{i,t}=\mu_i \mathrm{d} t+\sigma_{i,t} \mathrm{d}{W_{t}},
\end{equation}
where $\mu_i$ is the drift, $\sigma_{i,t}$ is the instantaneous volatility, and $W_t$ is the standard Brownian motion. The theoretical integrated variance (IV) of stock $i$ during $(t-h, t]$ is estimated as
\begin{equation}
\mathrm{IV}_{i,t}^{}(h)=\int_{t-h}^{t} \sigma_{i,s}^{2} \mathrm{d} s,
\end{equation}
where $h$ is the look-back horizon, such as 10 minutes, 30 minutes, 1 day, etc.

In this paper, we consider the minutely logarithmic mid-price return for asset $i$ during $(t-1, t]$ as
\begin{equation}
r_{i, t}:=\log \left(\frac{P_{i, t}}{P_{i, t-1}}\right).
\end{equation}
Here, $P_{i, t}$ is the mid-price at time $t$, i.e. $P_{i, t} = \frac{P_{i, t}^{b} + P_{i, t}^{s}}{2}$, and $P_{i, t}^{b}$ (respectively,  $P_{i, t}^{a}$) represents the best bid (respectively, ask) price. 

\citet{andersen2001distribution, barndorff2002econometric} showed that the sum of squared intraday returns is a consistent estimator of the unobservable IV. Because of the availability of high-frequency intraday data, we choose to compute realized volatility as a proxy for the unobserved IV (see \citet{bollen2002estimating, hansen2006realized, andersen2001distribution}). To reduce the impact of extreme values, we consider the logarithm, in line with \citet{andersen2003modeling, bucci2020realized, herskovic2016common}. Specifically, during a period $(t-h, t]$, the realized volatility is defined as follows\footnote{\citet{liu2015does} demonstrate that no sub-sampling frequency significantly outperforms a 5-min interval in terms of forecasting daily RVs, making it a widely accepted time interval in the literature. In the present paper, we use 1-min returns since our main focus is intraday RVs, such as 10-min RVs.} 

\vspace{-4mm}
\begin{equation}
\text{RV}_{i, t}^{(h)}:= \operatorname{log} \left[ \sum_{s=t-h+1}^{t} r_{i, s}^{2}\right].
\end{equation}

As pointed out by \citet{pascalau2021increasing}, there are no conclusive methods to incorporate the overnight session's information content into the daily volatility. In line with \citet{engle2012forecasting}, overnight information is excluded from our empirical analysis of daily volatility.
{For simplicity, we refer to this daily scenario (excluding the overnight) as the ``1-day'' scenario, throughout the rest of this paper.}

\subsection{Summary statistics}

\begin{figure}[H]
\centering
\subfigure[10-min]{\includegraphics[width=.4\textwidth, trim=0cm 0cm 1cm 1cm,clip]{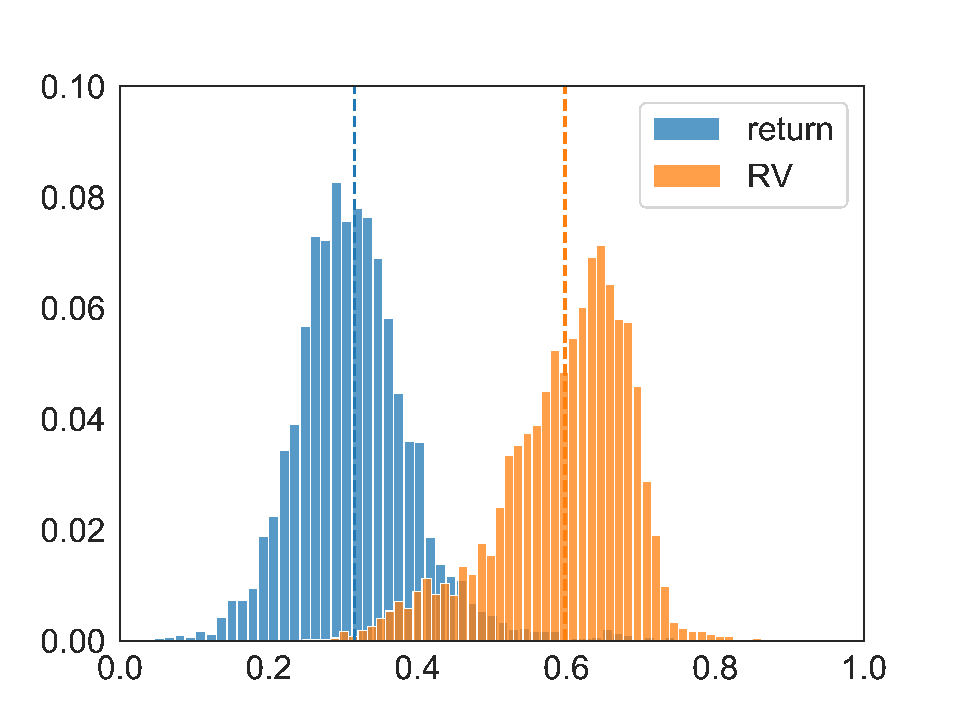}}
\subfigure[30-min]{\includegraphics[width=.4\textwidth, trim=0cm 0cm 1cm 1cm,clip]{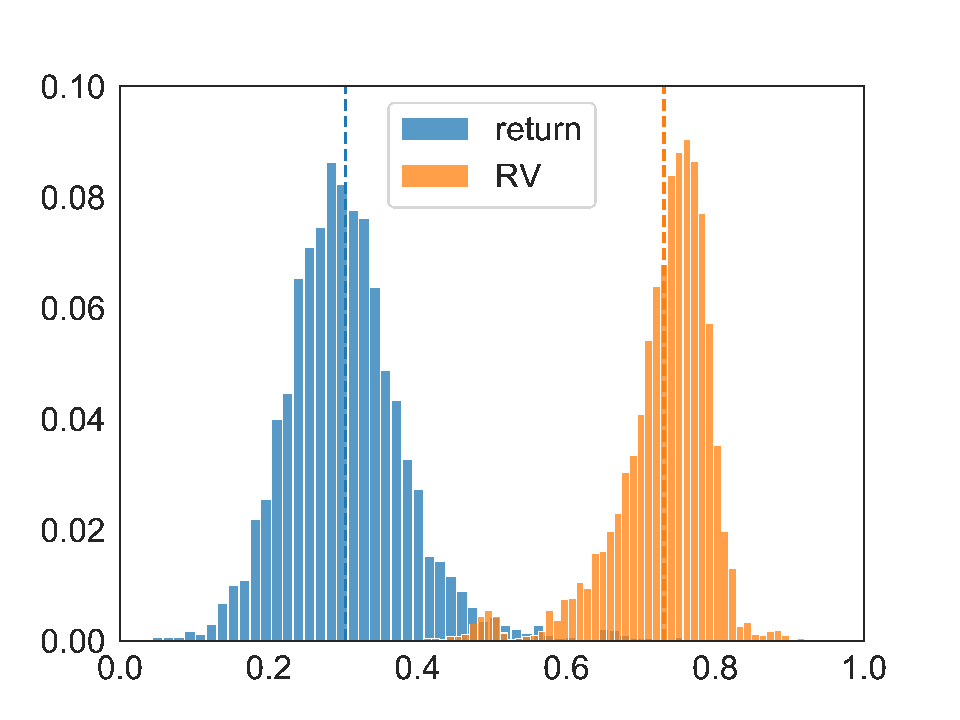}}
\subfigure[65-min]{\includegraphics[width=.4\textwidth, trim=0cm 0cm 1cm 1cm,clip]{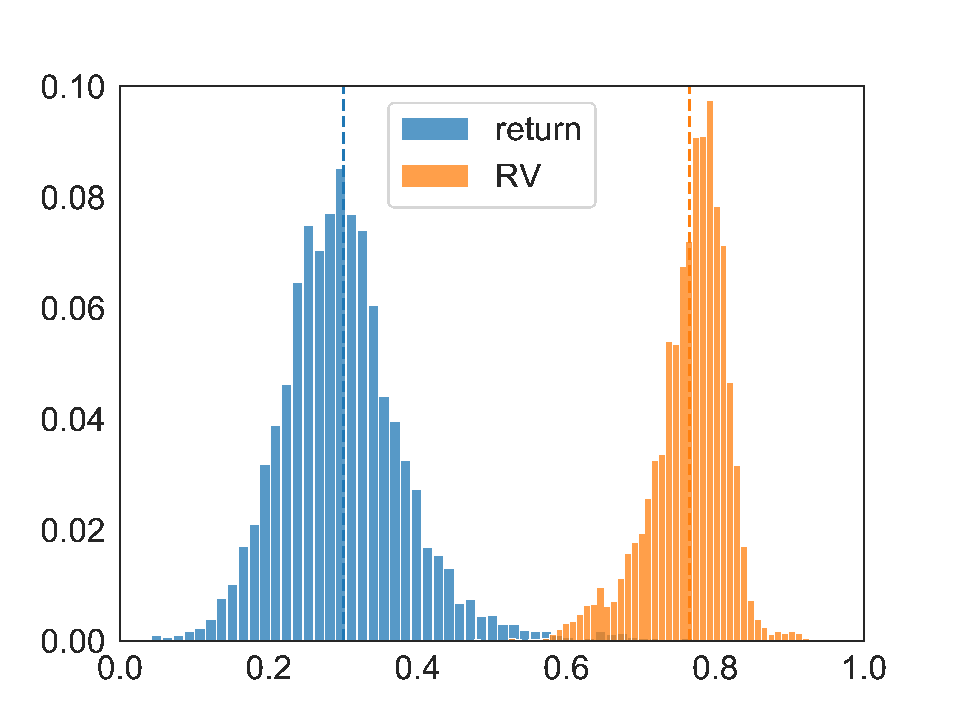}}
\subfigure[1-day]{\includegraphics[width=.4\textwidth, trim=0cm 0cm 1cm 1cm,clip]{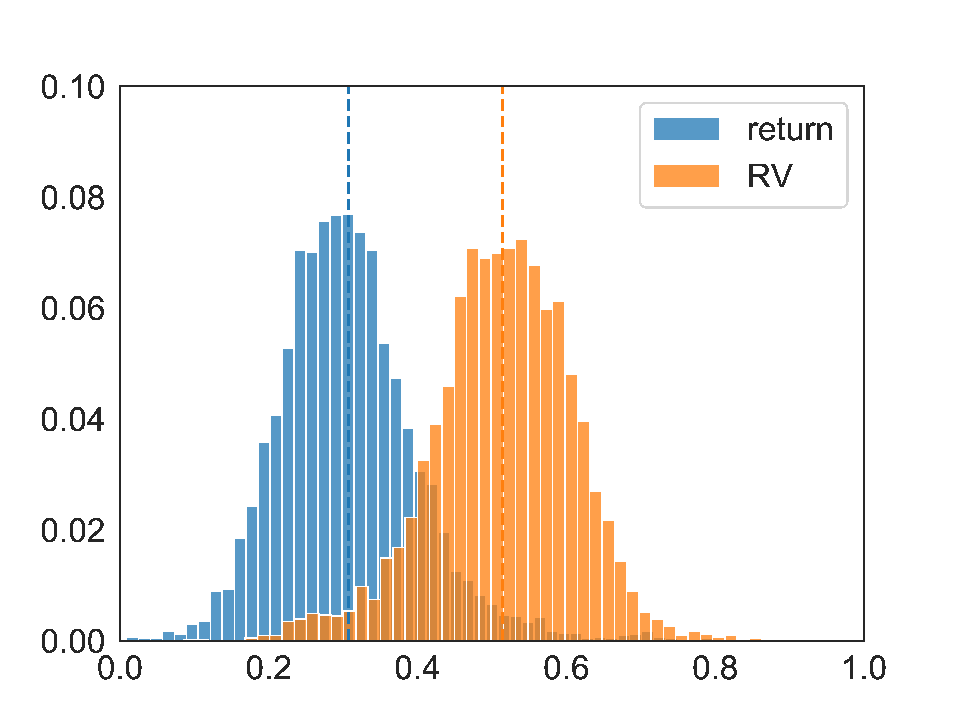}}
\caption{Histograms of pairwise correlations of realized volatilities and returns.}
\caption*{\textit{Notes:} (a)-(d) are based on observations in the frequency of 10-min, 30-min, 65-min, 1-day, respectively. The dashed vertical lines represent the average correlation values of RVs and returns.}
\vspace{-4mm} 
\label{fig:corr_daily_intra}
\end{figure}

To mitigate the effect of possibly spurious data errors, for each stock, {we set the values of return/volatility below the 0.5\% percentile equal to the respective 0.5\% percentile, and the values  above the 99.5\% percentile is set equal to the 99.5\% percentile,}  {a process commonly referred to as \textit{winsorization}}.  Figure \ref{fig:corr_daily_intra} illustrates the pairwise Pearson and Spearman correlations of returns and realized volatilities. This figure depicts the empirical distribution of pairwise correlation coefficients over the entire sample period. We generally observe  higher correlations in realized volatility than the counterparts in return. Figure  \ref{fig:corr_daily_intra} also reveals that, on average, as the horizon gets longer, realized volatility's correlations increase from 0.598 (10-min) to 0.731 (30-min) to 0.766 (65-min). However, when turning to daily realized volatility, correlations in RVs become weaker, with an average of 0.514. This indicates that the connections between stocks in terms of intraday volatility may be more stable and tight than the ones in daily volatility.

\begin{figure}[H]
\centering
\includegraphics[width=0.9\textwidth, trim=2.5cm 0mm 3cm  5mm,clip]{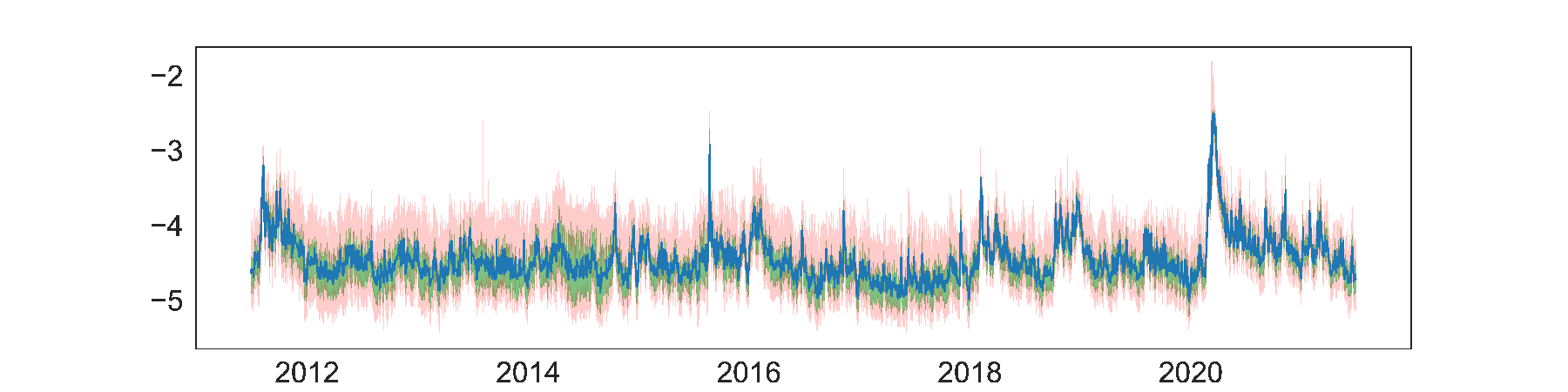}
\caption{Daily realized volatility (in logs). }
\caption*{\textit{Notes:} The blue curve represents cross-sectional average of daily realized volatility across stocks, with the green area covering the 25-th percentile to the 75-th percentile, and the red area covering the 5-th percentile to the 95-th percentile.}
\label{fig:daily_RV}
\end{figure}

\begin{figure}[H]
\centering
\includegraphics[width=0.9\textwidth, trim=2cm 0mm 3cm  5mm,clip]{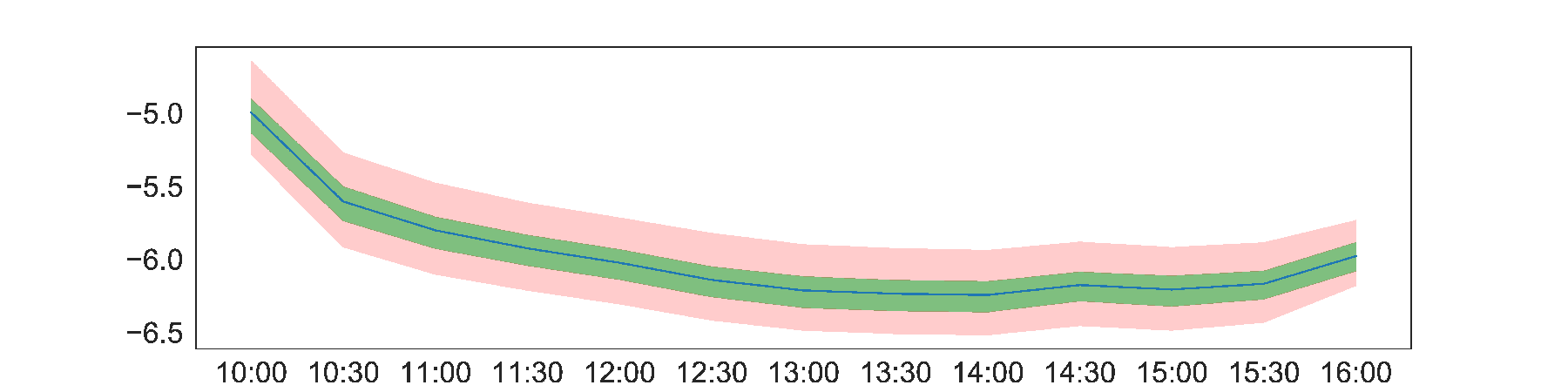}
\caption{Diurnal realized volatility (in logs).}
\caption*{\textit{Notes:} The blue curve represents cross-sectional average of 30-min realized volatility across stocks and days, with the green area covering the 25-th percentile to the 75-th percentile and the red area covering the  5-th percentile to the 95-th percentile.}
\label{fig:diurnal_RV}
\end{figure}

Figure \ref{fig:daily_RV} plots the daily realized volatility over time. Stocks demonstrate similar time series patterns, consistent with \citet{herskovic2016common, bollerslev2018risk}. Additionally, the width shrinks during the periods of higher volatility, such as, stock market crashes in August 2011 (European sovereign debt crisis), between June 2015 to June 2016 (Chinese stock market turbulence and Brexit), in March 2018 (China–United States trade war), in March 2020 (COVID-19). Figure \ref{fig:diurnal_RV} shows that the diurnal volatility forms a so-called reverse-J-shape, namely larger fluctuations near the open and close (see \citet{harris1986transaction, engle2012forecasting}).

\section{Commonality estimation}\label{sec:commonality}
Inspired by prior studies (e.g., \citet{morck2000information, chordia2000commonality, karolyi2012understanding, dang2015commonality}), we follow an analogous procedure to estimate the commonality in volatility. Specifically, we use the average adjusted R-squared value from the following regressions across stocks, as a measure of commonality in volatility (denoted as $R^{2}_{(h)}$)\footnote{We also perform another regression, where except for contemporaneous market volatility, {the lag one (thus $t-1$ in \eqref{eqn:common_RV}) and lead one (thus $t+1$ in \eqref{eqn:common_RV}, hence not computable in real time due to the forward looking bias)} in market volatility are also included, in order to explain non-contemporaneous trading, in line with \citet{chordia2000commonality, karolyi2012understanding, dang2015commonality}. The R-squared values are similar to the ones of Eqn \eqref{eqn:common_RV}.} 

\vspace{-4mm}
\begin{equation}\label{eqn:common_RV}
\text{RV}_{i, t}^{(h)} = \alpha_i + \beta_{i}\text{RV}_{M, t}^{(h)} +  \epsilon_{i,t},\\
\end{equation}
where $\text{RV}_{M, t}^{(h)}$ (see \citet{bollerslev2018risk}) is the contemporaneous market volatility during $(t-h, t]$ for stock $i$, which is calculated as the equally weighted average\footnote{We also implemented the value weighted market volatility and the results are similar to the equally weighted market volatility.} of all individual stock volatilities during $(t-h, t]$, i.e. 
\begin{equation}\label{eqn:mkt_RV}
\text{RV}_{M, t}^{(h)} = \frac{1}{N} \sum_{i=1}^N \text{RV}_{i, t}^{(h)}.
\end{equation}

\begin{figure}[H]
\centering
\includegraphics[width=0.75\textwidth, trim=1.4cm 5mm 3cm  1.5cm,clip]{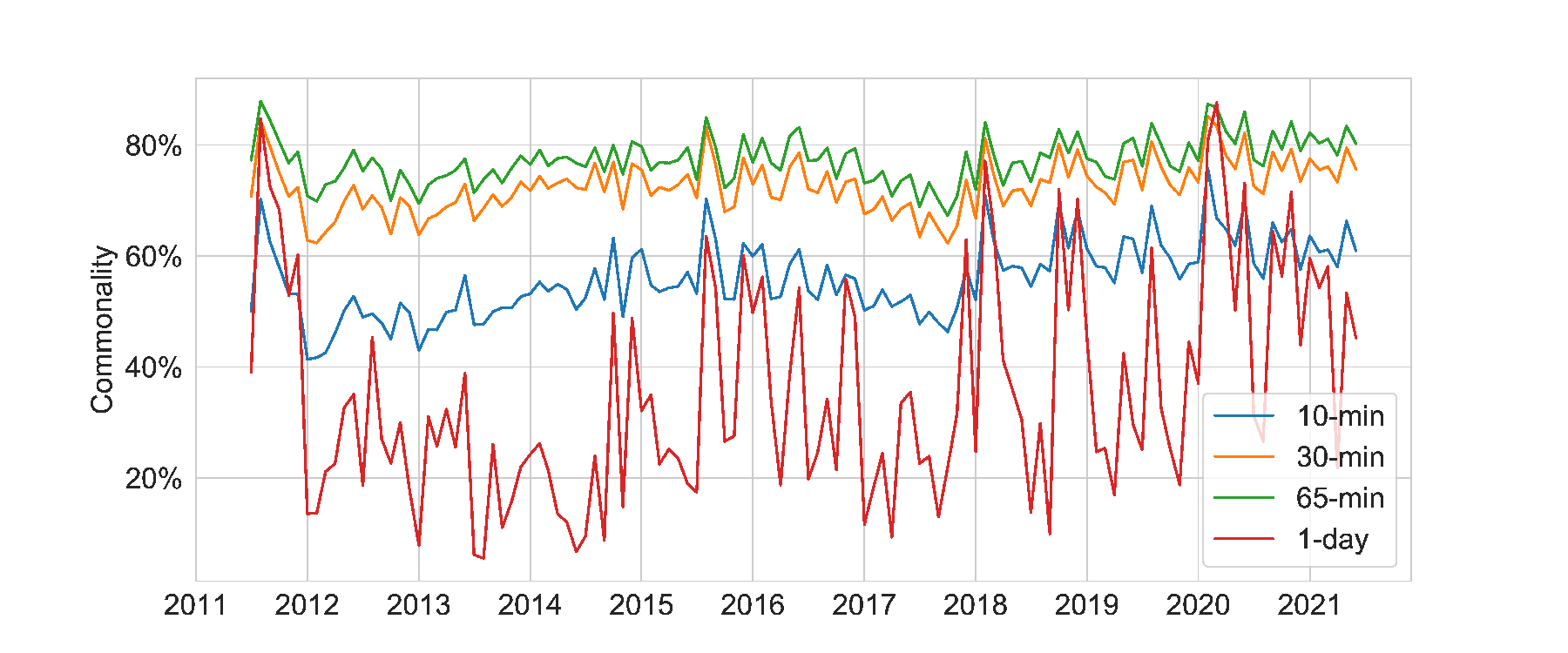}
\caption{Commonality in realized volatility.}
\caption*{\textit{Notes:} The commonality is averaged across stocks for each month during the sample period of 2011-07 $\sim$ 2021-06.}
\label{fig:common_ret_vol}
\end{figure}

Figure \ref{fig:common_ret_vol} presents the commonality in realized volatility, averaged across stocks for each month. To create this figure, we use the observations in each month, to obtain the R-squared value from Eqn \eqref{eqn:common_RV}. We notice that commonality effects in intraday scenarios (especially 30-min, 65-min) are substantially larger than the daily ones. {For example, as reported in  Table \ref{tab:common_stats}, the average commonality in 65-min data is around 74.3\%, while  only 35.5\% in daily data}. Moreover, $R^2_{(h)}$ is much more turbulent at the daily frequency. The last column in Table \ref{tab:common_stats} also reports the results of the relation between the average commonality and the market volatility. As the horizon extends, the average commonality has a higher correlation with the market volatility.\footnote{We refer the reader to additional analysis on commonality in Appendix \ref{app:common_factor}.}

\begin{table}[H]
\centering
\caption{Statistics of the monthly average commonality in volatility.}
\resizebox{0.42\textwidth}{!}{\begin{threeparttable}
\begin{tabular}{cccc}
   \toprule
   & Mean & Std   & Corr with VIX \\
   \midrule
10-min & 0.560 & 0.068 & 0.536 \\
30-min & 0.725 & 0.048 & 0.574 \\
65-min & 0.743 & 0.041 & 0.609 \\
1-day  & 0.355 & 0.198 & 0.690 \\
   \bottomrule
\end{tabular}
\end{threeparttable}}
\caption*{\textit{Notes:} VIX represents the market volatility from the Chicago Board Options Exchange.}
\label{tab:common_stats}
\end{table}

\begin{figure}[H]
\centering
\includegraphics[width=0.75\textwidth, trim=1.4cm 5mm 3cm 1.5cm,clip]{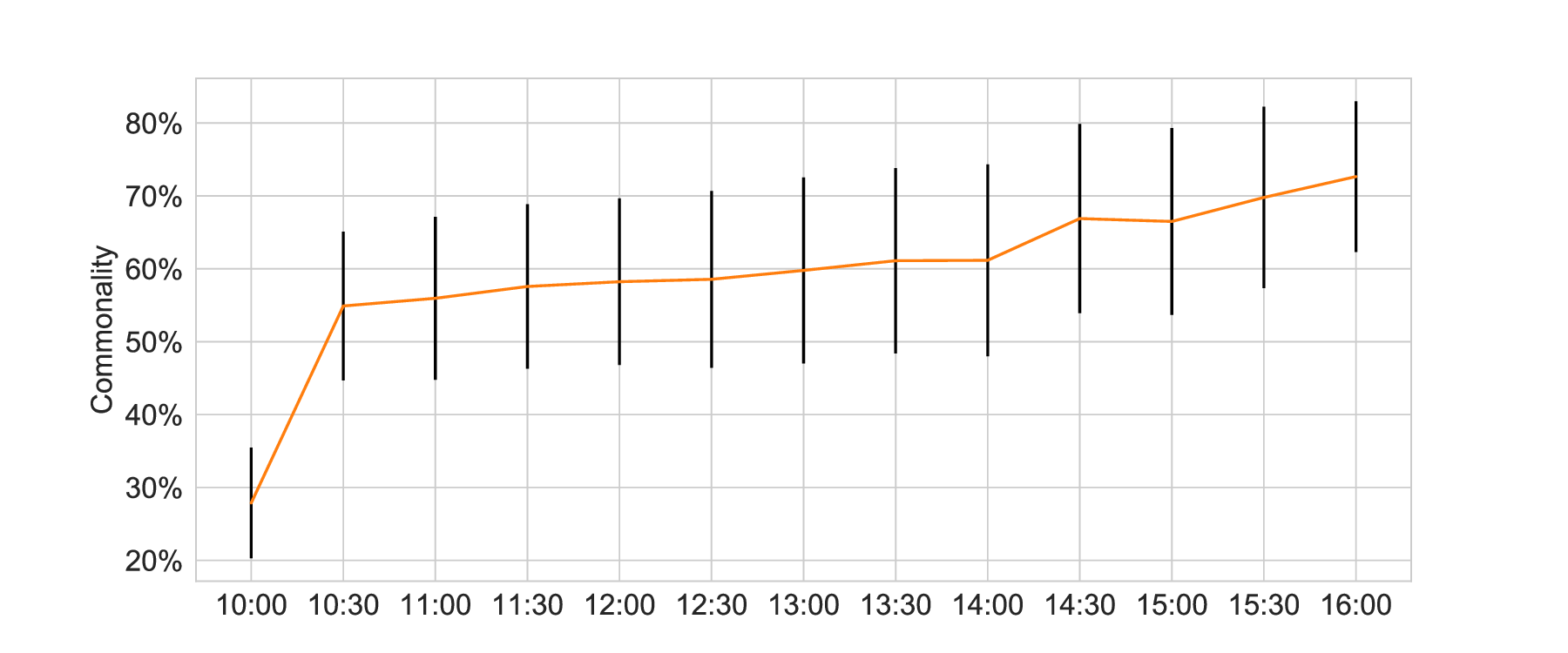}
\caption{Commonality in realized volatility.}
\caption*{\textit{Notes:} The commonality is averaged across stocks for each half-hour during the sample period of 2011-07 $\sim$ 2021-06.}
\label{fig:common_intra_vol}
\end{figure}

Figure \ref{fig:common_intra_vol} reports the averaged values and standard deviations (black vertical lines) of commonality for each half-hour in the trading session. To create this figure, we use the observations in a given interval, such as [09:30, 10:00], to fit Eqn \eqref{eqn:common_RV}. We observe a gradual increase in commonality throughout the trading session as we get closer to market close, {in sharp contrast to} the diurnal volatility pattern in Figure \ref{fig:diurnal_RV}. 

\section{Methodology} \label{sec:methodology}
In this section, we leverage the commonality for the task of predicting cross-asset volatility. We construct the prediction model as follows

\vspace{-1cm}
\begin{equation}\label{eq:general}
\begin{split}
    \text{RV}_{i,t+h}^{(h)} &= {F}_i\left(\mathbf{u}; \theta\right) + \epsilon_{i,t+h}\\
    &= {F}_i\left(\text{RV}_{i,t}^{(h)}, \dots, \text{RV}_{i,t-(p-1)h}^{(h)}, \text{RV}_{M, t}^{(h)}, \dots, \text{RV}_{M, t-(p-1)h}^{(h)} ; \theta \right) + \epsilon_{i,t+h}, 
\end{split} 
\end{equation}  
where $\text{RV}_{i,t+h}^{(h)}$ is the volatility of asset $i$ during $(t, t+h]$. $\mathbf{u}$ represents the input features, which can be further separated into 2 categories: (1) a multi-dimensional vector of predictor variables for a specific stock $i$ available up to time $t$, denoted as \textit{individual features}, such as $\left(\text{RV}_{i,t}^{(h)}, \dots, \text{RV}_{i,t-(p-1)h}^{(h)}\right)^{\prime};$ (2) a vector of features for all stocks in the studied universe up to $t$, denoted as \textit{market features}, such as $\left(\text{RV}_{M, t}^{(h)}, \dots, \text{RV}_{M, t-(p-1)h}^{(h)}\right)^{\prime}$. $\theta$ refers to the parameters that need to be estimated. {Whenever is clear from the context and no ambiguity arises, we use also use $\theta$ to denote the forecasting model.} 
We are aiming to find a function of variables that minimizes the out-of-sample errors for future realized volatility.

\subsection{Models} \label{sec:models}

This section summarizes the collection of machine learning models employed in our numerical experiments. 

\subsubsection{{Seasonal Autoregressive Integrated Moving Averages (SARIMA)}}
The Autoregressive Integrated Moving Averages (ARIMA) model is a popular forecasting method for univariate time series data, where an initial differencing step can be applied one or more times to eliminate the non-stationarity of the trend.
An ARIMA$(p, d, q)$ is given by
\begin{equation}\label{eq:arima}
\varphi(L)(1-L)^d\mathrm{RV}_{i, t}^{(h)}=\rho(L) \varepsilon_{i,t},
\end{equation}
where $\varphi(L)=1-\sum_{k=1}^p \varphi_i L^i$ and $\rho(L)=1-\sum_{j=1}^q \rho_j L^j$ are the AR and MA lag polynomials, and $\epsilon_{i,t}$ is the error which is distributed as $\mathcal{N}(0, \sigma_i^2)$. Following \citet{christensen1998relation, ribeiro2021novel}, we adopt ARIMA$(1, 1, 1)$ to model the daily realized volatility. 

When the time series exhibits seasonality, the seasonal-differencing could be applied to eliminate the seasonal component, which is denoted as Seasonal ARIMA (SARIMA). As revealed in Figure \ref{fig:diurnal_RV}, intraday volatility time series possess a seasonal component. For modeling intraday RV, we choose the SARIMA, where the seasonal period is the corresponding number of intraday time buckets in a day and other parameters related to the seasonal pattern are set as zero (for more details about SARIMA, see \citet{sheppard2010financial}).

\subsubsection{Heterogeneous autoregressive with diurnal effects (HAR-D)} \label{subsec:HAR}
\citet{corsi2009simple} proposed a volatility model, named as Heterogeneous autoregressive (HAR), which considers realized volatilities over different interval sizes. HAR has shown remarkably good forecasting performance on daily data \citet{patton2015good, izzeldin2019forecasting}. For day $t$, the forecast of HAR is based on
\begin{equation}\label{eq:har}
\mathrm{RV}_{i, t+1}^{(d)} = \alpha_i + \beta_i^{(d)} {\mathrm{RV}}_{i, t}^{(d)}+\beta_i^{(w)} {\mathrm{RV}}_{i, t}^{(w)}+ \beta_i^{(m)} {\mathrm{RV}}_{i, t}^{(m)}+\epsilon_{i, t+1},
\end{equation}
where ${\mathrm{RV}}_{i, t}^{(d)}$ denotes the daily realized volatility in the past day, and $\mathrm{RV}_{i, t}^{(w)} = \frac{1}{5} \sum_{l=1}^{5} \mathrm{RV}_{i, t-l}^{(d)},\quad  \mathrm{RV}_{i, t}^{(m)} = \frac{1}{21} \sum_{l=1}^{21} \mathrm{RV}_{i, t-l}^{(d)}$
denote the weekly and monthly lagged realized volatility, respectively. The choice of a daily, weekly and monthly lag is aiming to capture the long-memory dynamic dependencies observed in most realized volatility series.

However, very little attention has been paid to forecasting intraday volatility with HAR. One closely connected model is that of \citet{engle2012forecasting}, who  proposed an intraday volatility forecasting model, where they interpret that conditional volatility of high-frequency returns is a product of daily, diurnal, and stochastic intraday components. After the decomposition of raw returns, the authors apply a GARCH model \citet{engle1982autoregressive} to learn the stochastic intraday volatility components.

Following the spirit of \citet{engle2012forecasting}, we extend the daily HAR model to intraday scenarios by adding diurnal effect and previous intraday component, as follows\footnote{Since we use the log-version realized volatility, the multiplication of daily, diurnal, and stochastic intraday components in \citet{engle2012forecasting} translates to the addition in our model \eqref{eq:har_diurnal}.}     

\vspace{-5mm}
\begin{equation}\label{eq:har_diurnal}
\mathrm{RV}_{i, t+h}^{(h)} = \alpha_i + \beta_{i}^{(\tau)} {D}_{i, \tau_{t+h}} + \beta_{i}^{(s)} \mathrm{RV}_{i, t}^{(h)} +\beta_i^{(d)} {\mathrm{RV}}_{i, t}^{(d)}+\beta_i^{(w)} {\mathrm{RV}}_{i, t}^{(w)}+ \beta_i^{(m)} {\mathrm{RV}}_{i, t}^{(m)}+\epsilon_{i, t+h},
\end{equation}
where ${D}_{i, \tau_{t+h}}$ denote the average diurnal realized volatility in the bucket-of-the-day $\tau_{t+h}$ computed from the last 21 days. For example, when $t=$ 10:30 and $h=30$ minutes, then $\tau_{t+h}$ corresponds to the bucket 10:30-11:00. $\mathrm{RV}_{i, t}^{(h)}$ represents the lag=1 intraday RV. ${\mathrm{RV}}_{i, t}^{(d)}$ (${\mathrm{RV}}_{i, t}^{(w)}$, ${\mathrm{RV}}_{i, t}^{(m)}$) denotes the aggregated daily (weekly, monthly) realized volatility. 
When we consider the daily scenarios, Eqn \eqref{eq:har_diurnal} becomes the standard HAR model (Eqn \eqref{eq:har}), by removing the diurnal term and the intraday component. 

\subsubsection{Ordinary least squares (OLS)}\label{subsec:ols}
Instead of using aggregated realized volatility, we apply OLS to original features, as follows, with its loss function being the sum of squared errors. Recall $\mathbf{u}=(u_1, \dots, u_p)^{\prime}$ represent the vector of input features, such as past intraday RVs, and (perhaps) market RVs. Notice that the model only incorporating the past intraday RVs as features is actually an autoregressive (AR) model.

\vspace{-3mm}
\begin{equation}\label{eq:ols}
\mathrm{RV}_{i, t+h}^{(h)} = \alpha_i + \sum_{l=1}^p \beta_{l} u_l +\epsilon_{i, t+h}.
\end{equation}

\subsubsection{Least absolute shrinkage and selection operator (LASSO)} 
When the number of predictors approaches the number of observations, or there are high correlations among predictor variables, the OLS model tends to overfit noise rather than signals. This is particularly burdensome for the volatility forecasting problem, where the features could be highly correlated. 

LASSO is a linear regression method that can avoid overfitting via adding a penalty of parameters to the objective function. As pointed out by \citet{hastie2009elements}, LASSO performs both variable selection and regularization, therefore enhances the prediction accuracy and interpretability of regression models. The objective function of LASSO is the sum of squared residuals and an additional $l_1$ constraint on the regression coefficients, as shown in Eqn \eqref{eq:loss_lasso}. Here, the hyperparameter $\lambda$ controls the penalty weight. In our experiments, we provide a set of hyperparameter values, and then choose the one with the best performance on the validation data, as our forecasting model.

\vspace{-4mm}
\begin{equation}\label{eq:loss_lasso}
L_{i} = \sum_{t} \left[\mathrm{RV}_{i, t+h}^{(h)} - \alpha_i - \sum_{l=1}^p \beta_{l} u_l \right]^2 + \lambda \sum_{l=1}^p \left\|\beta_{l}\right\|_1.
\end{equation}

\subsubsection{XGBoost}
Linear models are unable to capture the possible non-linear relations between the dependent variable and the predictors, and the interactions among predictors. As pointed by \citet{bucci2020realized}, RVs are subject to structural breaks and regime-switching, hence the need to consider non-linear models. One way to add non-linearity and interactions is the decision tree, see more in \citet{hastie2009elements}.

\begin{figure}[H]
\centering
\includegraphics[width=.9\textwidth]{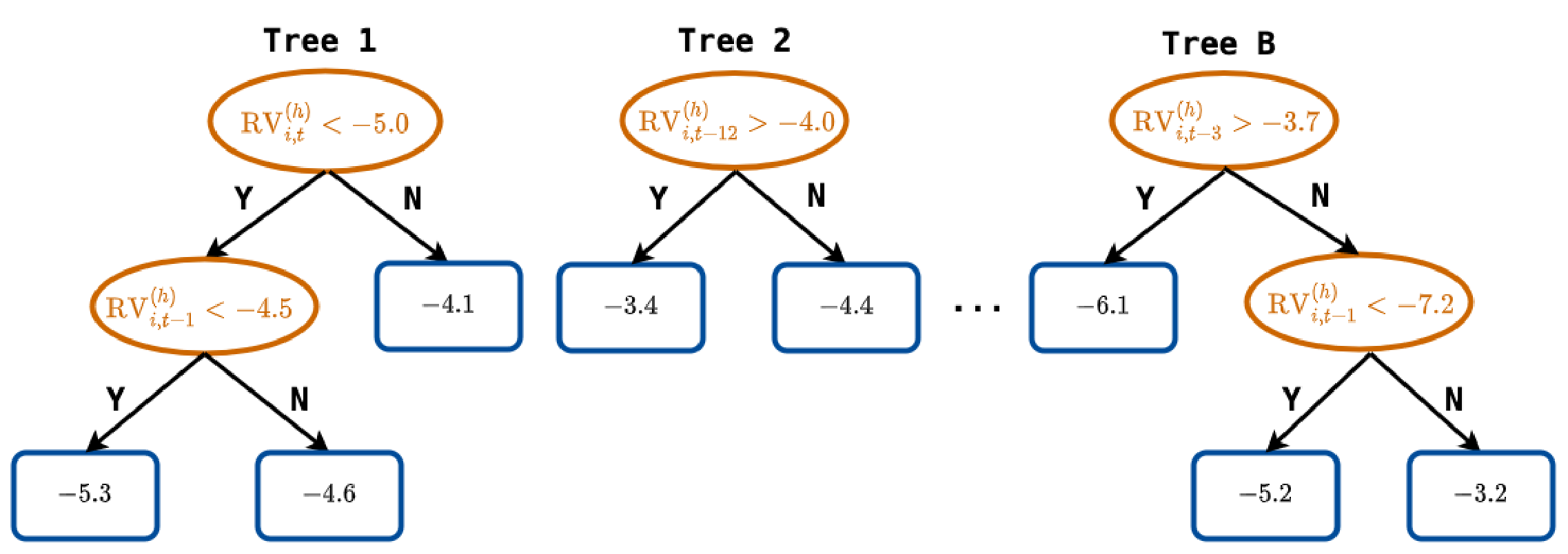}
\caption{Illustration of a tree ensemble model.}
\caption*{\textit{Notes:} B represents the number of trees. The final prediction of a tree ensemble model is the sum of predictions from each tree, as shown in Eqn \eqref{eq:ensemble_tree}.}
\label{fig:ensemble_tree}
\end{figure}

XGBoost is a decision-tree-based ensemble algorithm, implemented under a distributed gradient boosting framework by \citet{chen2016xgboost}. There is abundant empirical evidence showing the success of XGBoost, such as in a large number of Kaggle competitions. In this work, we only review the essential idea behind XGBoost - tree boosting model. 
For more details about other important features of XGBoost, such as the scalability in various scenarios, parallelization, distributed computing, {feature importance to enhance interpretability}, 
etc., the reader may refer to \citet{chen2016xgboost}. Let $\mathbf{u}$ represent the vector of input features,
\vspace{-3mm}
\begin{equation}\label{eq:ensemble_tree}
{F}_i\left(\mathbf{u}\right)=\sum_{l=1}^{B} f_{l}\left(\mathbf{u}\right), \quad f_{l} \in \mathcal{F},
\end{equation}
where $\mathcal{F}$ is the space of regression trees. An example of the tree ensemble model is depicted in Figure \ref{fig:ensemble_tree}. The tree ensemble model in Eqn \eqref{eq:ensemble_tree} is trained sequentially. Boosting (see \citet{friedman2001greedy}) means that new models are added to minimize the errors made by existing models, until no further improvements are achieved.

\subsubsection{Multilayer perceptron (MLP)}

{Another non-linear method is the neural network (NN), which has become increasingly popular for machine learning problems, e.g. in computer vision and natural language processing, due to its flexibility to learn complex interactions. \citet{hill1996neural} and \citet{zhang2003time} suggest that NN is a promising non-linear alternative to the traditional linear methods in time series forecasting. We  introduce two commonly-implemented NNs in the following sections.}

MLP is a class of feedforward neural networks and a ``universal approximator'' that can learn any smooth functions (see \citet{hornik1989multilayer}). MLP has been applied to many fields, e.g. computer vision, natural language processing, etc. MLPs are composed of an input layer to receive the raw features, an output layer that makes forecasts about the input, and in-between those two, an arbitrary number of hidden layers that are non-linear transformations. MLPs perform a static mapping between an input space and an output space \cite{bucci2020realized}. The parameters in MLPs can be updated via stochastic gradient descent (SGD). In this work, we use Adam (see \citet{kingma2014adam}), which is based on adaptive estimates of lower-order moments. Let $\mathbf{u} \in \mathbb{R}^{p} $ represent the input variables
\vspace{-3mm}
\begin{equation}\label{eq:mlp}
{F}_i\left(\mathbf{u}; \mathbf{\theta} \right) = \mathbf{W}_L \cdot \sigma \left( \mathbf{W}_{L-1} \ldots \sigma (\mathbf{W}_1 \mathbf{u} +\mathbf{b}_1)\ldots +\mathbf{b}_{L-1}\right) +\mathbf{b}_L,
\end{equation}
where $\mathbf{\theta}:=(\mathbf{W}_1,\mathbf{W}_2,\ldots,\mathbf{W}_L,\mathbf{b}_1,\mathbf{b}_2,\ldots,\mathbf{b}_L)$ represents the parameters in the neural network. $\mathbf{W}_l \in \mathbb{R}^{n_l \times n_{l-1}}$, $\mathbf{b}_l \in \mathbb{R}^{n_l \times 1}$ for $l=1,2,\ldots,L$, and $n_{0}=p$. For the activation function $\sigma(\cdot)$, we choose the rectified linear unit (ReLU), i.e. $\sigma(x) = \operatorname{max}(x, 0)$.

\subsubsection{Long short-term memory (LSTM)} 
{A recurrent neural network (RNN) requires that historic information is retained to forecast future values. Therefore RNN is well-suited for processing, classifying, and making predictions based on time series data \cite{bucci2020realized}. LSTM, proposed by \citet{hochreiter1997long}, is an extension of the RNN architecture by replacing each hidden unit in RNNs with a memory block to capture the long-term effect. LSTMs have received considerable success in natural language processing, time series, generative models, and etc. }

For simplicity, we consider the time series for a given stock and remove the subscript for stock identity. The standard transformation in each unit of LSTM is defined as follows. For a more detailed discussion, we refer the reader to  \citet{hochreiter1997long}. 

\vspace{-5mm}
\begin{equation}
\begin{aligned}
\mathbf{f}_{t} &=\sigma_{g}\left(\mathbf{W}_{f} \mathbf{u}_{t}+\mathbf{U}_{f} \mathbf{h}_{t-1}+\mathbf{b}_{f}\right) \\
\mathbf{i}_{t} &=\sigma_{g}\left(\mathbf{W}_{i} \mathbf{u}_{t}+\mathbf{U}_{i} \mathbf{h}_{t-1}+\mathbf{b}_{i}\right) \\
\mathbf{o}_{t} &=\sigma_{g}\left(\mathbf{W}_{o} \mathbf{u}_{t}+\mathbf{U}_{o} \mathbf{h}_{t-1}+\mathbf{b}_{o}\right) \\
\tilde{\mathbf{c}}_{t} &=\sigma_{c}\left(\mathbf{W}_{c} \mathbf{u}_{t}+U_{c} \mathbf{h}_{t-1}+\mathbf{b}_{c}\right) \\
\mathbf{c}_{t} &= \mathbf{f}_{t} \circ \mathbf{c}_{t-1}+\mathbf{i}_{t} \circ \tilde{\mathbf{c}}_{t} \\
\mathbf{h}_{t} &=\mathbf{o}_{t} \circ \sigma_{h}\left(\mathbf{c}_{t}\right),
\end{aligned}
\end{equation}
where $\mathbf{u}_{t}$ is input vector, $\mathbf{f}_{t}$ is forget gate's activation vector, $\mathbf{i}_{t}$ is update gate's activation vector, $\mathbf{o}_{t}$ is output gate's activation vector, $\tilde{\mathbf{c}}_{t}$ is cell input activation vector, $\mathbf{c}_{t}$ is cell state vector, and $\mathbf{h}_{t}$ is hidden state vector, i.e. output vector of the LSTM unit. $ \circ$ is the Hadamard product function. $\sigma_{g}$ is sigmoid function, and $\sigma_{c}, \sigma_{h}$ are hyperbolic tangent function.  $\mathbf{W}_{f(i, o, c)}, \mathbf{b}_{f(i,o,c)}$ refer to weight matrices and bias vectors that need to be estimated. 

To summarize, we first consider a traditional time series model ARIMA, then include three linear regression models, i.e. HAR(-D), OLS, and LASSO. To account for the nonlinear impact of individual predictors on future volatilities and the interactions among predictors, we choose an ensemble tree model XGBoost and two neural networks, i.e. MLP and LSTM. The primary difference between MLP and LSTM is that LSTM has feedback connections, which allow learning the dependencies in the input sequences.

\subsection{Training scheme}
Motivated by the strong commonality in volatility across stocks, we consider the following three different schemes for model training. 
\begin{itemize}
    \item {\single}  denotes that we train customized models $F_i$ for each stock $i$, as in \citet{bucci2020realized} and \citet{hansen2005forecast}. We use a stock's own past RVs only as predictor features, namely 
    \[\mathbf{u} =  \left(\text{RV}_{i,t}^{(h)}, \dots, \text{RV}_{i,t-(p-1)h}^{(h)}\right)^{\prime}
    \]
    and no market features, where $p$ represents the number of lags. 

    \item {\univ} denotes that we train models with the pooled data of all stocks in our universe. {That is}, $F_i$ is same for all stocks in Eqn \eqref{eq:general}. As in the {\single} scheme, we use a stock's own past RVs only as predictor features and no market features. \citet{sirignano2019universal} showed that the model trained on the pooled data outperforms asset-specific models trained on time series of any given stock,  in the sense of forecasting the direction of price moves. \citet{bollerslev2018risk,engle2012forecasting} suggested that models estimated under the {\univ} setting yield superior out-of-sample risk forecasts, compared to models under the {\single} setting,  when forecasting daily realized volatility.
    
    \item {\aug} denotes that we train models with the pooled data of all stocks in our universe, but in addition, we also incorporate a predictor which takes into account the impact of the market realized volatility (e.g. \citet{bollerslev2018risk}) in order to leverage the commonality in volatility shown in Section  \ref{sec:commonality}. Namely, $F_i$ is same for all stocks in Eqn \eqref{eq:general}. We use both individual features and market features as predictors, i.e. $\mathbf{u} = \left(\text{RV}_{i,t}^{(h)}, \dots, \text{RV}_{i,t-(p-1)h}^{(h)}, \text{RV}_{M, t}^{(h)}, \dots, \text{RV}_{M, t-(p-1)h}^{(h)}\right)^{\prime}$. Note that for HAR-D models under the {\aug} setting, we include aggregated market features as additional features,  and use the OLS  to estimate the parameters.
\end{itemize}

In summary, compared to the benchmark {\single} setting, we  gradually incorporate cross-asset and market information into the training of models. The hyperparameters for each model are summarized in Appendix \ref{app:hyperparameter}.

\subsection{Performance evaluation} \label{sec:evaluation}
To assess the predictive performance for RV forecasts, we compute the following metrics on the rolling out-of-sample tests (see \citet{patton2009evaluating, patton2011volatility, engle2012forecasting, pascalau2021increasing, bollerslev2018risk, bucci2020realized, rahimikia2020machine}). Both functions measure losses, so lower values are preferred. \citet{patton2009evaluating} demonstrate that QLIKE has the highest power in the Diebold–Mariano (DM) test. {Consequently, we focus more on the QLIKE rather than the MSE.}
\begin{itemize}
    \item {Mean squared error (MSE)}: $\frac{1}{N} \sum_{i=1}^{N} \frac{1}{ \#\mathcal{T}_{test}} \sum_{t\in \mathcal{T}_{test}}\left(\text{RV}_{i, t}^{(h)}-\widehat{\text{RV}}_{i, t}^{(h)}\right)^{2},$
    \item {Quasi-likelihood (QLIKE)}: $\frac{1}{N} \sum_{i=1}^{N} \frac{1}{\#\mathcal{T}_{test}} \sum_{t\in \mathcal{T}_{test}} \left[ \frac{\exp(\text{RV}_{i, t}^{(h)})}{\exp(\widehat{\text{RV}}_{i, t}^{(h)})} - (\text{RV}_{i, t}^{(h)}-{\widehat{\text{RV}}_{i, t}^{(h)}})-1 \right],$
\end{itemize}
where $\widehat{\text{RV}}_{i, t}^{(h)}$ represents the predicted value of $\text{RV}_{i, t}^{(h)}$, the realized volatility for stock $i$ during $(t-h, t]$, and $\overline{\text{RV}}_{i}^{(h)}$ is the empirical mean of $\text{RV}_{i, t}^{(h)}$ on the test data. $N$ is the number of stocks in our universe, $\mathcal{T}_{test}$ is the testing period, and $\#\mathcal{T}_{test}$ is the length of the testing period. 

\paragraph{{Model Confidence Set (MCS)}} was proposed by \cite{hansen2011model} to identify a subset of models $\mathcal{M}^*$ with significantly superior performance from model candidates $\mathcal{M}_0$, at a given level of confidence. The iterative elimination is based on sequentially testing the following hypothesis 

\vspace{-3mm}
\begin{equation}
 H_0: \mathbb{E}\left(\Delta L_{ij, t}\right)=0, \mbox{for all } i, j \in \mathcal{M}^*, 
\end{equation}  

\noindent   where $L_{ij, t}$ is the loss difference between models $i$ and $j$ at day $t$ in terms of a specific loss function $L$, such as MSE and QLIKE. The MCS procedure renders it possible to make statements about the statistical significance from multiple pairwise comparisons. For additional details, we refer to the studies of \cite{hansen2011model}.

\subsection{{Utility benefits}}
{We have demonstrated that one can assess the out-of-sample statistical performance for each model via the above metrics and tests. However, in such an approach, the economic magnitude of the gain from complex risk models is ignored. \citet{bollerslev2018risk} have proposed a utility-based framework, which gauges the utility benefits of an investor with mean-variance preferences investing in an asset with time-varying volatility and a constant Sharpe ratio. We implement this framework to measure the volatility forecasts. For a more detailed description, we refer the reader to \citet{bollerslev2018risk}.}

{The expected utility of a mean-variance investor at $t$ can be approximated as
\begin{equation}\label{eqn:mv_1}
\mathbb{E}_{t}\left(u\left(W_{t+1}\right)\right)=\mathbb{E}_{t}\left(W_{t+1}\right)-\frac{1}{2} \gamma^{A} \operatorname{Var}_{t}\left(W_{t+1}\right),
\end{equation}
where $W_{t}$ denotes the wealth and $\gamma^{A}$ denotes the absolute risk aversion of the investor. }

{Assume that the investor allocates a fraction $x_t$ of the current wealth to a risky asset with return $r_{t+1}$ and the rest to a risk-free money market account earning $r_{t}^f$. Then the wealth at $t+1$ becomes $W_{t+1}=W_{t}\left(1+r_{t}^{f}+x_{t} r_{t+1}^{e}\right)$, where $r_{t+1}^{e} \equiv r_{t+1}-r_{t}^{f}$. After dropping constant terms, the expected utility in Eqn \eqref{eqn:mv_1}  amounts to   
\begin{equation}\label{eqn:mv_util}
\mathbb{E}_{t}\left(u\left(W_{t+1}\right)\right):=U\left(x_{t}\right) =W_{t}\left(x_{t} \mathbb{E}_{t}\left(r_{t+1}^{e}\right)-\frac{\gamma}{2} x_{t}^{2} \mathbb{E}_{t}\left(\exp(\text{RV}_{t+1})\right)\right),
\end{equation}
where $\gamma \equiv \gamma^{A} W_{t}$ denotes the investor’s relative risk aversion.}

Suppose the conditional Sharpe ratio $SR \equiv \mathbb{E}_{t}\left(r_{t+1}^{e}\right) / \sqrt{\mathbb{E}_{t}\left(\exp(\text{RV}_{t+1})\right)}$ is constant, so that the expected utility is 
\begin{equation}\label{eqn:mv_2}
U\left(x_{t}\right)=W_{t}\left(x_{t} SR \sqrt{\mathbb{E}_{t}\left(\exp(\text{RV}_{t+1})\right)}-\frac{\gamma}{2} x_{t}^{2} \mathbb{E}_{t}\left(\exp(\text{RV}_{t+1})\right)\right).
\end{equation}

The optimal portfolio that maximizes this utility is obtained by investing the following fraction of wealth to the risky asset
\begin{equation}\label{eqn:opt}
x_{t}^{*}=\frac{SR / \gamma}{\sqrt{\mathbb{E}_{t}\left(\exp(\text{RV}_{t+1})\right)}}.
\end{equation}

To determine the utility gains based on different risk models,  the expectation based on model $\theta$ is denoted by $\mathbb{E}_{t}^{\theta}(\cdot)$. Assuming that the investor uses model $\theta$, then the position $x_{t}^{\theta}=\frac{SR / \gamma}{\sqrt{\mathbb{E}_{t}^{\theta}\left(\exp(\text{RV}_{t+1})\right)}}$ is chosen. By plugging $x_{t}^{\theta}$ into Eqn \eqref{eqn:mv_2} and replacing $\mathbb{E}_{t}(\exp(\text{RV}_{t+1}))$ with the realized volatility $\exp(\text{RV}_{t+1})$, the expected utility per unit of the wealth (called realized utility, or in short RU) is given by 
\begin{equation}\label{eqn:uti}
\mathrm{RU}_t = \frac{SR^2}{\gamma} \times \frac{\sqrt{\exp(\text{RV}_{t+1})}}{\sqrt{\mathbb{E}_{t}^{\theta}(\exp(\text{RV}_{t+1}))}} - \frac{SR^2}{2\gamma} \times \frac{\exp(\text{RV}_{t+1})}{\mathbb{E}_{t}^{\theta}(\exp(\text{RV}_{t+1}))}.
\end{equation}

If a risk model is ideal, that is, it predicts perfectly the realized volatilities $\mathbb{E}_{t}^{\theta}(\exp(\text{RV}_{t+1})) = \exp(\text{RV}_{t+1})$, then its realized utility is  $\frac{SR^2}{2\gamma}$. Alternatively, the investor is willing to give up $\frac{SR^2}{2\gamma}$ of the wealth in order to utilize the perfect risk model instead of investing only in the risk-free asset. In this paper,  the same Sharpe ratio (SR = 0.4)  and the same coefficient of risk aversion ($\gamma=2$) are applied as in \citet{bollerslev2018risk, li2020forecasting}.

The previous comparisons are based on a frictionless setting, ignoring the trading cost. The case of incorporating the effect of transaction costs is also considered. Following \citet{bollerslev2018risk} and \citet{li2020forecasting}, we assume that transaction costs are linear in the absolute magnitude of the change in the positions, and use the full median bid-ask spread for each of the assets over the last 90 trading days. The realized utility with trading costs deducted, denoted as RU-TC,  is simply the realized utility after subtracting the simulated costs. We evaluate this realized utility (with and without  trading cost) empirically by averaging the corresponding realized expressions over stocks and the same rolling out-of-sample forecasts.

\section{Experiments} \label{sec:exp}

\subsection{Implementation} \label{sec:setup}
For each data set, we divide the observations into three non-overlapping periods and maintain their chronological order: training, validation, and testing. For a given trading day $t$, the training data, including the samples in the first period [2011-07-01$, t-251]$, are used to estimate models subject to a given architecture. Validation data, including the recent one-year samples $[t-250, t]$, are deployed to tune the hyperparameters of the models. Finally, testing data are samples in the next year $[t+1, t+251]$; they are out-of-sample in order to provide objective assessments of the models' performance. Due to limited computational resources, models are updated annually. In other words, when we retrain the models in the next calendar year, the training data expands by one year, whereas the validation samples are rolled forward to include the samples in the most recent one-year period, following \citet{gu2020empirical}. {To examine the effect of model update frequency, we perform a robustness check for HAR-D models in Appendix \ref{app:model_update}, and we conclude that the update frequency has insignificant effect on the model's performance.} Our testing period  starts from 2015-07-01 until 2016-06-30, and the corresponding  training and validation samples are [2011-07-01, 2014-06-30] and [2014-07-01, 2015-06-30], respectively. When we predict the realized volatility in [2016-07-01, 2017-06-30], the training and validation samples are [2011-07-01, 2015-06-30] and [2015-07-01, 2016-06-30], respectively. Therefore, our testing sample includes 6 years, from July 2015 to June 2021. 

For HAR-D and OLS, we use both the training and validation data for training, due to no requirement of hyperparameter tuning. Given the stochastic nature of neural networks, we apply an ensemble approach to MLPs and LSTMs for improving their robustness (see \citet{hansen1990neural, gu2020empirical}). Specifically, we train multiple neural networks with different random seeds for initialization, and construct final predictions by averaging the outputs of all networks. For more information on the model settings, see Appendix \ref{app:hyperparameter}.

In all of the models, the features are based on the observations in the last 21 days. Prior to inputting variables in the models, at each rolling window estimation, we  normalize them by removing the mean and scaling to unit variance. 

\subsection{Main results} \label{sec:results}
Table \ref{tab:experiment_sigmetaaug} presents the results of each model under three training schemes.\footnote{To more formally assess the statistical significance of the differences in out-of-sample volatility forecasts, Table \ref{tab:DM} in Appendix \ref{app:DM} also reports the results of all Diebold-Mariano tests in terms of QLIKE.} Due to limited computation power, MLPs and LSTMs are only performed under the {\univ} and {\aug} settings. We draw the following conclusions from Table \ref{tab:experiment_sigmetaaug}.

We begin by comparing the SARIMA with HAR-D under the {\single} setting.\footnote{Note that (S)ARIMA is for single time series.} The results show that the ARIMA model achieves similar performance as HAR over the 1-day horizon, consistent with \citet{izzeldin2019forecasting}. However, HAR-D yields more accurate out-of-sample forecasts than SARIMA across intraday horizons, i.e. 10-min, 30-min, 65-min. 

Regarding linear models, we observe that for HAR-D, {\univ} shows no improvement in forecasting, compared to {\single}. HAR-D models trained under {\aug} significantly outperform the ones trained under the other two schemes, across all horizons in our study. The average reduction in QLIKE of {\aug} compared to {\single} is 0.031, -0.005, 0.004, 0.010 over 10-min, 30-min, 65-min, and 1-day, respectively.

\begin{table}[H]
\centering
\caption{Out-of-sample performance.}
\resizebox{\textwidth}{!}{\begin{tabular}{l l c c c c c c c c }
    \toprule
  \multicolumn{2}{c}{\textbf{ Panel A:} }  & \multicolumn{2}{c}{{10-min}} & \multicolumn{2}{c}{{30-min}} & \multicolumn{2}{c}{{65-min}} & \multicolumn{2}{c}{{1-day}} \\
     \cmidrule(lr){3-4}\cmidrule(lr){5-6}\cmidrule(lr){7-8}\cmidrule(lr){9-10}
     
 \multicolumn{2}{c}{\textbf{ Statistical performance} } & MSE & QLIKE & MSE & QLIKE & MSE & QLIKE & MSE & QLIKE \\ \midrule
{SARIMA} & {\single}   & 1.319 & 0.617 & 0.524 & 0.338 & 0.362 & 0.231 & 0.268 &  0.190 \\\midrule
&{\single}   &1.013 &0.484 &0.332 &0.222 &0.270 &0.190 &0.269 &0.190  \\
                        &{\univ}  &1.021 &0.518 &0.333 &0.230 &0.270 &0.191 &0.269 &0.190 \\
                    \multirow{-3}{*}{HAR-D}    &{\aug}  &0.995 &0.453 &0.323 &0.227 &0.262 &0.186 &0.257 &0.180  \\ \midrule
 & {\single} & 1.009 & 0.490 & 0.307 & 0.222 & 0.251 & 0.176 & 0.263 & 0.192 \\
 & {\univ} & 1.008 & 0.507 & 0.307 & 0.221 & 0.250 & 0.175 & 0.260 & 0.191 \\
\multirow{-3}{*}{OLS} & {\aug} & 0.962 & 0.430 & 0.293 & 0.204 & 0.241 & 0.171 & \textcolor{red}{0.254}* & \textcolor{red}{0.178}* \\ \midrule
 & {\single} & 1.053 & 0.492 & 0.325 & 0.224 & 0.252 & 0.180 & 0.263 & 0.195 \\
 & {\univ} & 1.012 & 0.511 & 0.309 & 0.222 & 0.251 & 0.176 & 0.261 & 0.192 \\
\multirow{-3}{*}{LASSO} & {\aug} & 0.961 & 0.428 & 0.293 & 0.204 & 0.242 & 0.172 & \textcolor{blue}{0.255}* & \textcolor{blue}{0.179}* \\ \midrule
 & {\single} & 1.047 & 0.539 & 0.345 & 0.236 & 0.297 & 0.201 & 0.358 & 0.217 \\
 & {\univ} & 0.968 & 0.417 & 0.290 & 0.191 & 0.242 & 0.170 & 0.268 & 0.192 \\
\multirow{-3}{*}{XGBoost} & {\aug} & 0.968 & 0.422 & 0.297 & 0.190 & 0.249 & 0.174 & 0.285 & 0.187 \\ \midrule
 & {\single} & - & - & - & - & - & - & - & - \\
 & {\univ} & 0.947 & 0.397 & 0.284 & 0.181 & 0.232 & 0.163 & 0.260 & 0.191 \\
\multirow{-3}{*}{MLP} & {\aug} & \textcolor{blue}{0.945} & \textcolor{blue}{0.386} & \textcolor{blue}{0.280}* & \textcolor{blue}{0.179} & \textcolor{blue}{0.229}* & \textcolor{blue}{0.162} & 0.257 & 0.180 \\ \midrule
 & {\single} & - & - & - & - & - & - & - & - \\
 & {\univ} & 0.950 & 0.393 & 0.287 & 0.179 & 0.232 & 0.162 & 0.261 & 0.188 \\
\multirow{-3}{*}{LSTM} & {\aug} & \textcolor{red}{0.934}* & \textcolor{red}{0.376}* & \textcolor{red}{0.279}* & \textcolor{red}{0.171}* & \textcolor{red}{0.229}* & \textcolor{red}{0.160}* & 0.258 & 0.182\\

\midrule
  \multicolumn{2}{c}{\textbf{ Panel B:} }  & \multicolumn{2}{c}{{10-min}} & \multicolumn{2}{c}{{30-min}} & \multicolumn{2}{c}{{65-min}} & \multicolumn{2}{c}{{1-day}} \\
     \cmidrule(lr){3-4}\cmidrule(lr){5-6}\cmidrule(lr){7-8}\cmidrule(lr){9-10}
     
 \multicolumn{2}{c}{\textbf{  Realized utility} } & RU & RU-TC & RU & RU-TC & RU & RU-TC & RU & RU-TC\\ \midrule 
{SARIMA} &{\single}   & 2.095 & 1.473 & 3.041 & 2.624 & 3.314 & 2.861 & 3.550 & 3.515  \\\midrule
&{\single}   &2.690 &2.065 &3.457 &3.040 &3.542 &3.095 &3.548 &3.515  \\
                        &{\univ}  &2.574 &1.975 &3.429 &3.016 &3.541 &3.095 &3.547 &3.514\\
                        \multirow{-3}{*}{{HAR-D}}&{\aug}  &2.790 &2.280 &3.428 &3.022 &3.552 &3.107 &3.571 &3.536  \\ \midrule
 & { {\single}} & { 2.660} & { 1.901} & { 3.506} & { 3.027} & { 3.579} & { 3.118} & { 3.543} & { 3.504} \\
& { {\univ}} & { 2.601} & { 2.039} & { 3.432} & { 2.984} & { 3.580} & { 3.127} & { 3.546} & { 3.513} \\
\multirow{-3}{*}{{ OLS}} & { {\aug}} & { {2.845}} & { 2.271} & { 3.485} & { 3.036} & { 3.587} & { 3.130} & { \textcolor{red}{3.576}} & { \textcolor{red}{3.536}} \\\midrule
 & { {\single}} & { 2.631} & { 1.893} & { 3.526} & { 3.061} & { 3.567} & { 3.108} & { 3.545} & { 3.501} \\
 & { {\univ}} & { 2.593} & { 2.044} & { 3.432} & { 2.989} & { 3.578} & { 3.126} & { 3.543} & { 3.512} \\
\multirow{-3}{*}{{ LASSO}} & { {\aug}} & { 2.852} & { 2.292} & { 3.487} & { 3.046} & { 3.586} & { 3.132} & { \textcolor{blue}{3.575}} & { \textcolor{blue}{3.537}} \\\midrule
 & { {\single}} & { 2.492} & { 1.552} & { 3.408} & { 2.888} & { 3.520} & { 3.039} & { 3.508} & { 3.449} \\
 & { {\univ}} & { 2.890} & { 2.200} & { 3.532} & { 3.067} & { 3.592} & { 3.116} & { 3.546} & { 3.505} \\
\multirow{-3}{*}{{ XGBoost}} & { {\aug}} & { 2.864} & { 2.212} & { 3.545} & { 3.083} & { 3.581} & { 3.109} & { 3.571} & { 3.524} \\\midrule
 & { {\single}} & { -} & { -} & { -} & { -} & { -} & { -} & { -} & { -} \\
{ } & { {\univ}} & { 2.952} & { 2.380} & { 3.564} & { 3.119} & { 3.607} & { 3.139} & { 3.543} & { 3.506} \\
\multirow{-3}{*}{{ MLP}} & { {\aug}} & { \textcolor{blue}{2.993}} & { \textcolor{blue}{2.442}} & { {3.569}} & { {3.126}} & { 3.609} & { 3.145} & { 3.571} & { 3.534} \\\midrule
 & { {\single}} & { -} & { -} & { -} & { -} & { -} & { -} & { -} & { -} \\
 & { {\univ}} & { 2.975} & { 2.455} & { \textcolor{blue}{3.575}} & { \textcolor{blue}{3.144}} & { \textcolor{blue}{3.610}} & { \textcolor{blue}{3.149}} & { 3.552} & { 3.514} \\
\multirow{-3}{*}{{ LSTM}} & { {\aug}} & { \textcolor{red}{3.028}} & { \textcolor{red}{2.532}} & { \textcolor{red}{3.595}} & { \textcolor{red}{3.170}} & { \textcolor{red}{3.614}} & { \textcolor{red}{3.166}} & { 3.567} & { 3.533}\\
\bottomrule
\end{tabular}}
\caption*{\textit{Notes:} The table reports the out-of-sample results for predicting future realized volatility over multiple horizons using different models under three training schemes. For each horizon, the model with the best (second best) out-of-sample performance in QLIKE (in Panel A) / RU (in Panel B) is highlighted in \textcolor{red}{red} (\textcolor{blue}{blue}), respectively. An asterisk (*) indicates models that are included in the MCS at the 5\% significance level.}
\label{tab:experiment_sigmetaaug}
\end{table}

Generally speaking, there are significant improvements when moving from HAR-D models to OLS models, over 10-min, 30-min, and 65-min horizons. For example, QLIKEs are reduced from 0.453 (resp. 0.227, 0.186) with the best HAR-D model (i.e. under {\aug}) to 0.430 (resp. 0.204, 0.171)  with the best OLS model (i.e. under {\aug}), across the three horizons (i.e. 10, 30, 65-min), respectively. Within the OLS models, conclusions are similar with HAR-D models, i.e. no benefits from {\univ} while significant benefits from {\aug}. We also observe similar findings in LASSO as in OLS, {suggesting that regularization does not further aid performance}. On the other hand, MLPs and LSTMs achieve state-of-the-art accuracy across all measures and intraday horizons (i.e. 10, 30, 65-min), implying the complex interactions between predictors. Further analysis is provided in Section \ref{sec:interaction}. 
    
Interestingly, linear models slightly outperform MLPs and LSTMs at the 1-day horizon. {This is perhaps expected, and might be due to the availability of only a small amount of data at the 1-day horizon, rendering the neural networks to underperform due to lack of training data}. 

{Echoing the findings from Panel A, OLS based on the 21-day rolling daily RVs deliver the higher utility than the HAR type models, consistent with \citet{bollerslev2018risk}. Neural networks still perform the best, with the highest realized utility achieved by LSTMs.}

\begin{figure}[H]
\centering
\subfigure[OLS, {10-min}\label{fig:diff_ts_10m}]{\includegraphics[width=.47\textwidth, trim=0mm 0mm 2cm 1cm, clip]{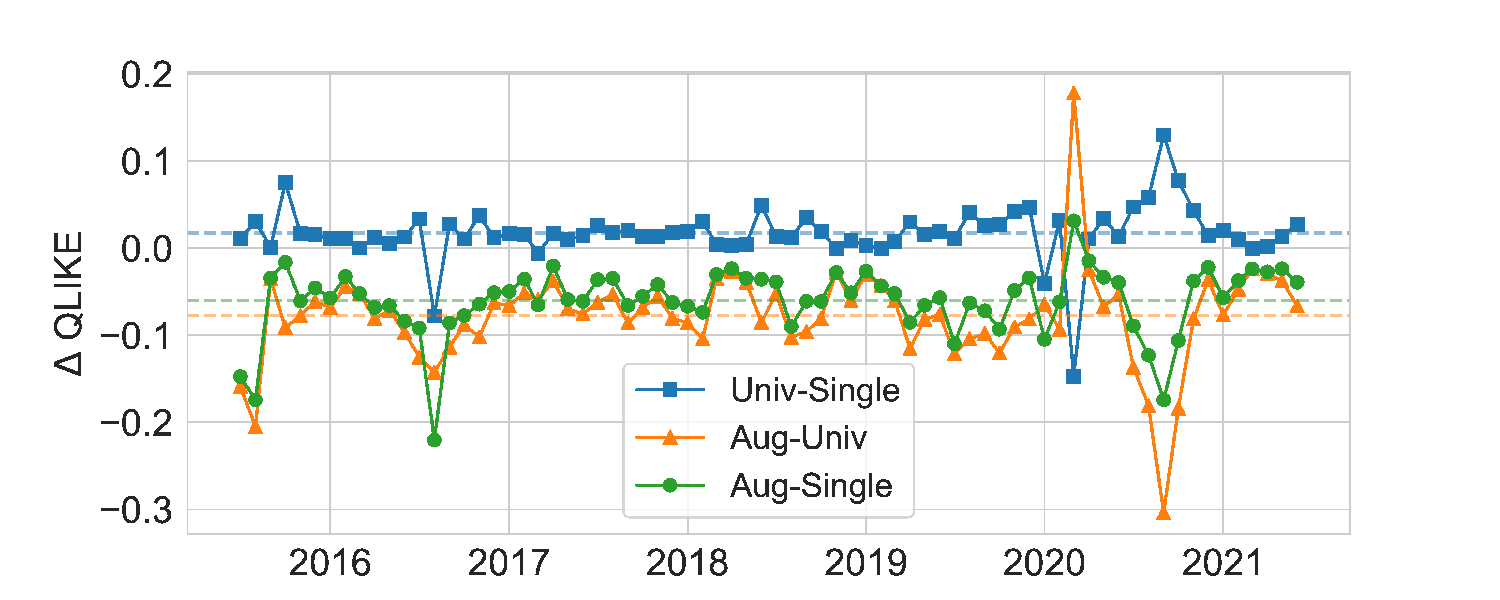}}
\subfigure[OLS, 30-min\label{fig:diff_ts_30m}]{\includegraphics[width=.47\textwidth, trim=0mm 0mm 2cm 1cm,clip]{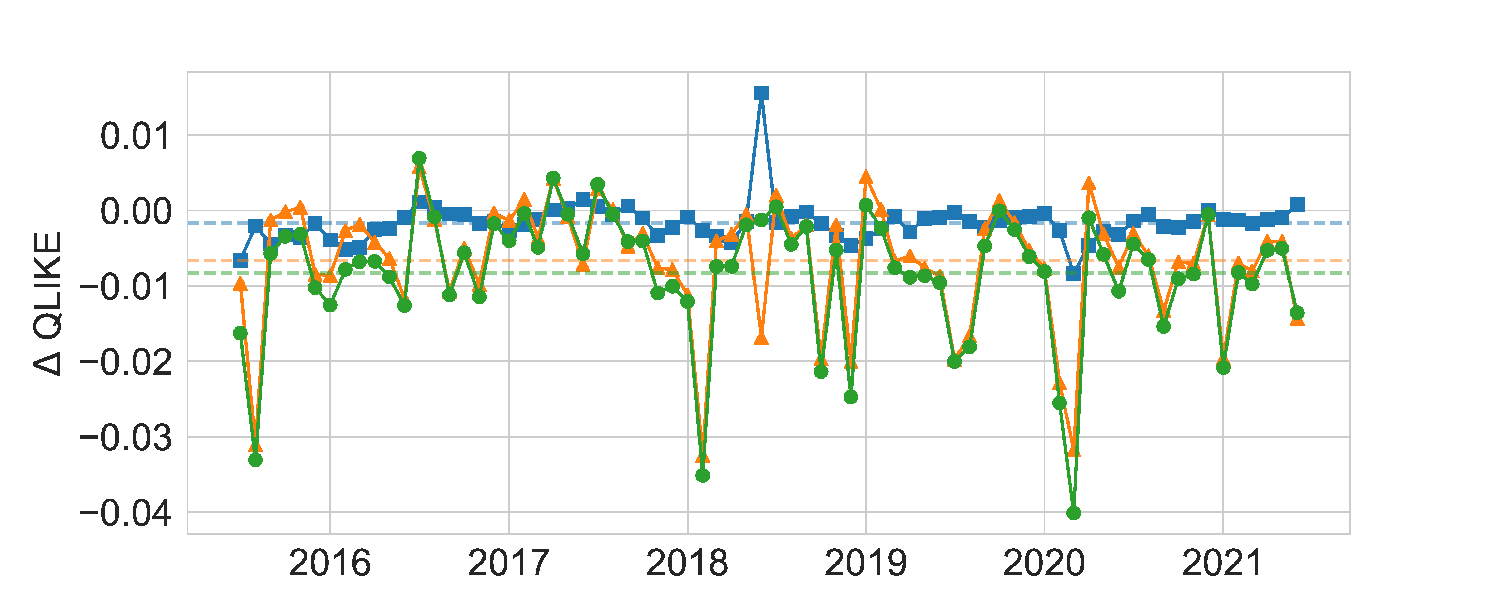}}
\subfigure[OLS, 65-min\label{fig:diff_ts_65m}]{\includegraphics[width=.47\textwidth, trim=0mm 0mm 2cm 1cm, clip]{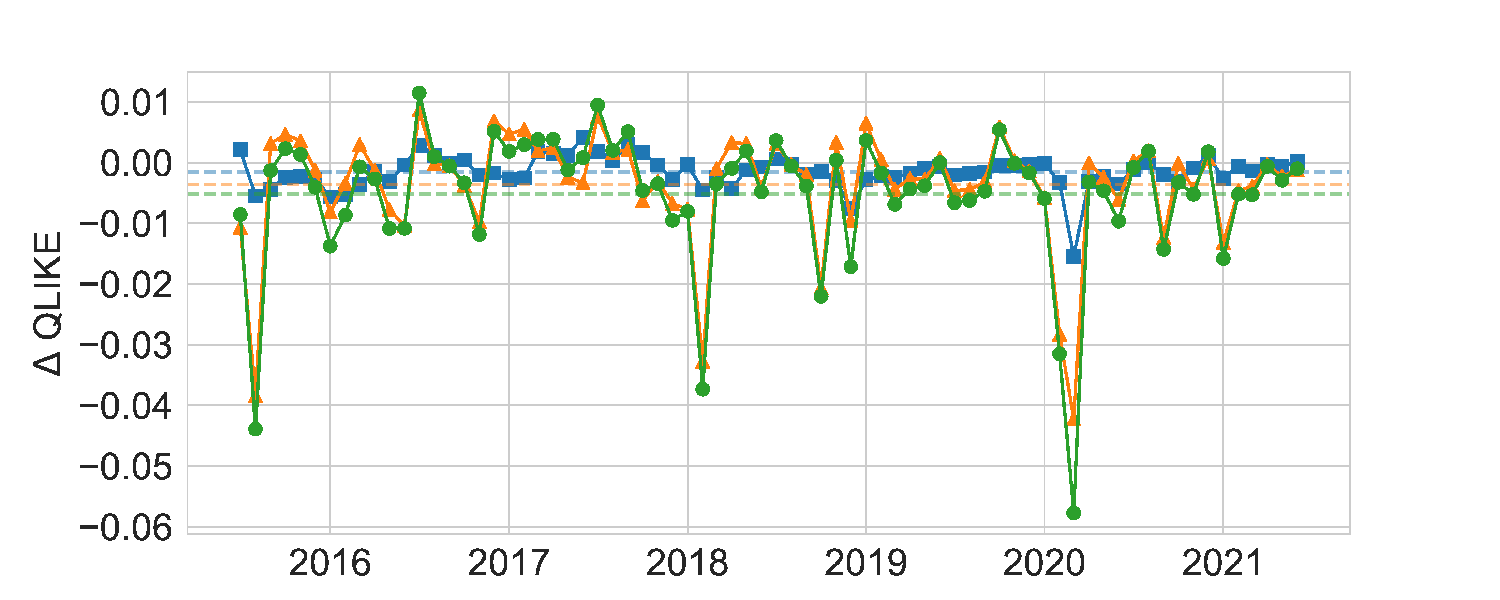}}
\subfigure[OLS, 1-day\label{fig:diff_ts_1d}]{\includegraphics[width=.47\textwidth, trim=0mm 0mm 2cm 1cm,clip]{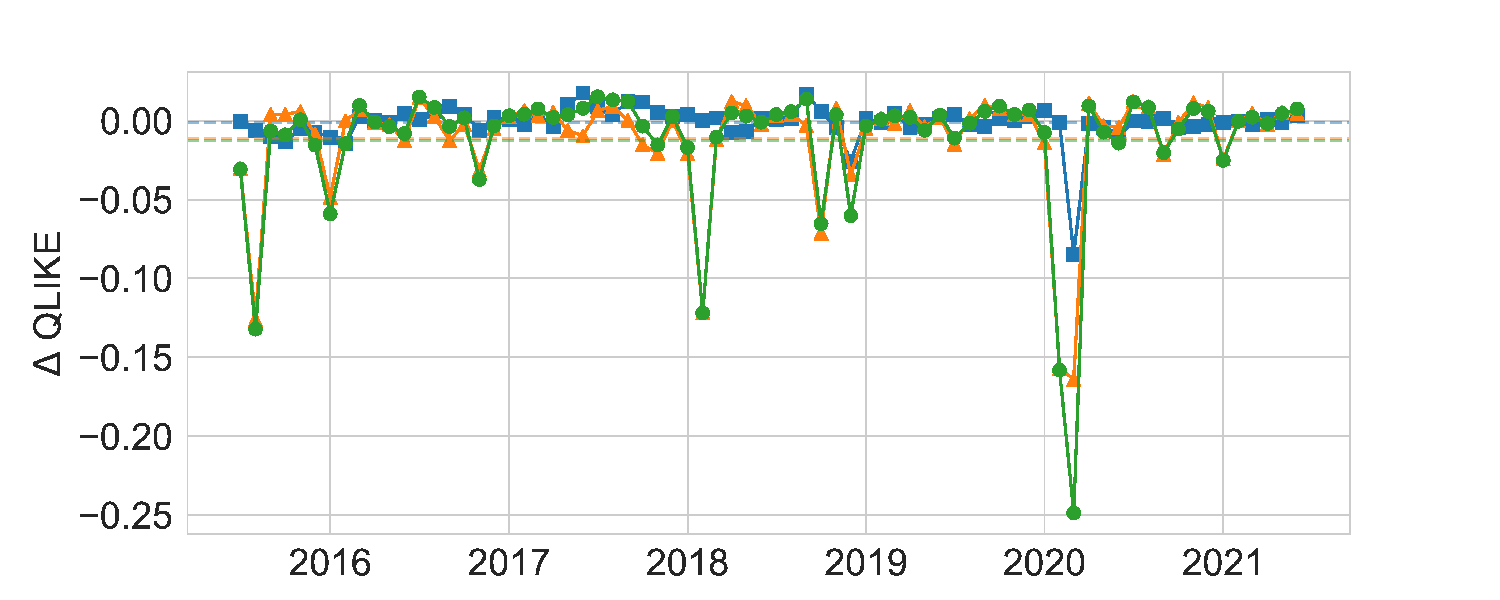}}
\caption{Pairwise $\Delta$QLIKE of the OLS model  across three training schemes.}
\caption*{\textit{Notes:} $\Delta$QLIKE is averaged across stocks in each month during the testing period 2015-07 $\sim$ 2021-06. The dashed horizontal lines represent the average reductions in QLIKE.}
\label{fig:diff_ts}
\end{figure}

Let us now consider the OLS model as an illustrative example for understanding the relative reduction in error.
We compare its QLIKEs under these three schemes, at a monthly level, as shown in Figure \ref{fig:diff_ts}. For better readability, we report the reduction in error of {\univ} relative to {\single} (denoted as Univ-Single), the reduction of {\aug} relative to {\univ} (denoted as Aug-Univ), and the reduction of {\aug} relative to {\single} (denoted as Aug-Single). Note that Aug-Single = (Aug-Univ) + (Univ-Single).
Negative values of $\Delta \text{QLIKE}$ indicate an improvement on out-of-sample data, and positive values indicate degradation. To arrive at this figure, we average the $\Delta \text{QLIKE}$ values in each month, across stocks. Figure \ref{fig:diff_ts} reveals that the improvement of {\univ} compared to {\single} is relatively small but consistent. In terms of the benefits of {\aug}, it is typically the case that incorporating the market volatility as an additional feature helps improve the forecasting performance, especially for turmoil periods.

\begin{figure}[H]
\centering
\subfigure[OLS, {10-min}\label{fig:diff_common_10min}]{\includegraphics[width=.47\textwidth, trim=0mm 0mm 1cm 1cm, clip]{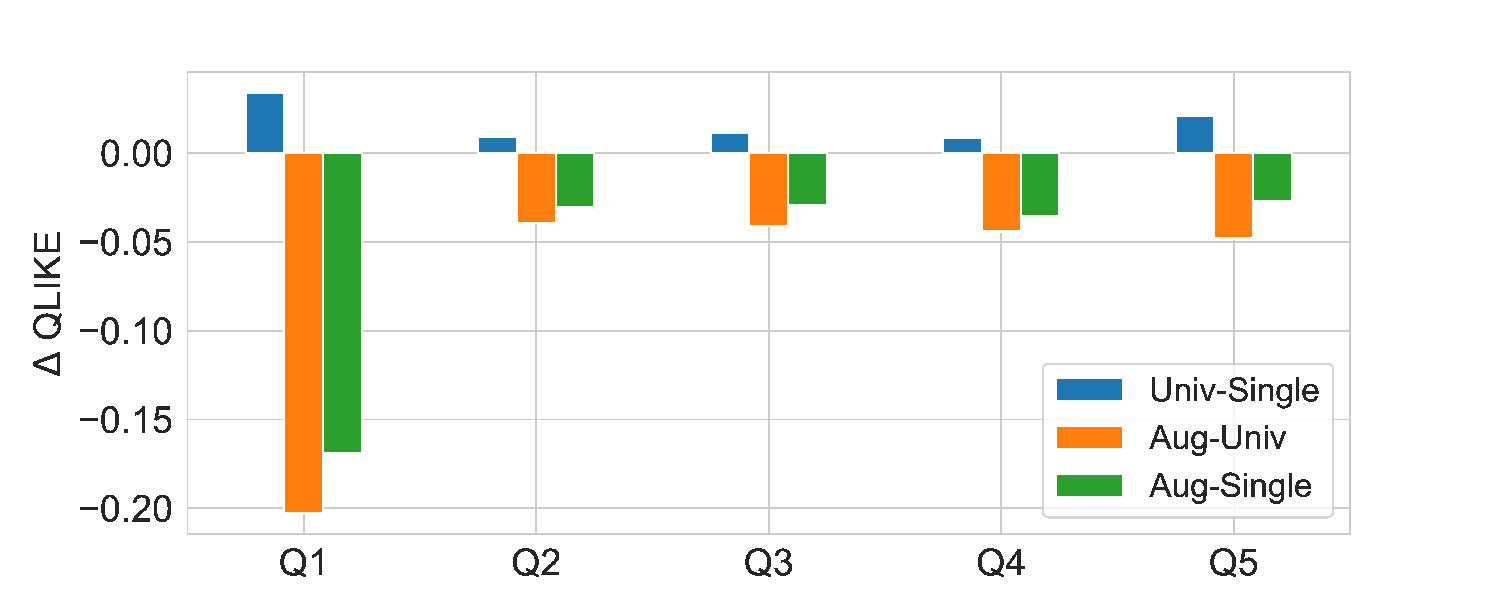}}
\subfigure[OLS, 30-min\label{fig:diff_common_30min}]{\includegraphics[width=.47\textwidth, trim=0mm 0mm 1cm 1cm, clip]{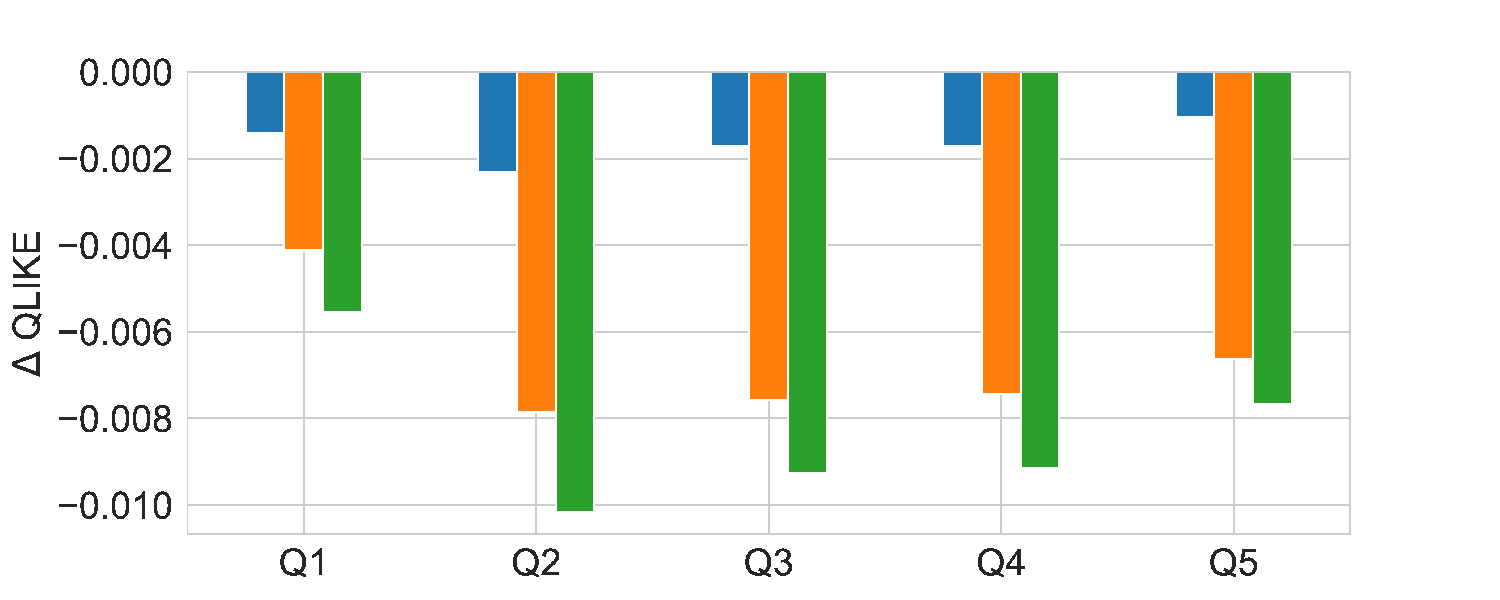}}
\subfigure[OLS, {65-min}\label{fig:diff_common_65min}]{\includegraphics[width=.47\textwidth, trim=0mm 0mm 1cm 1cm, clip]{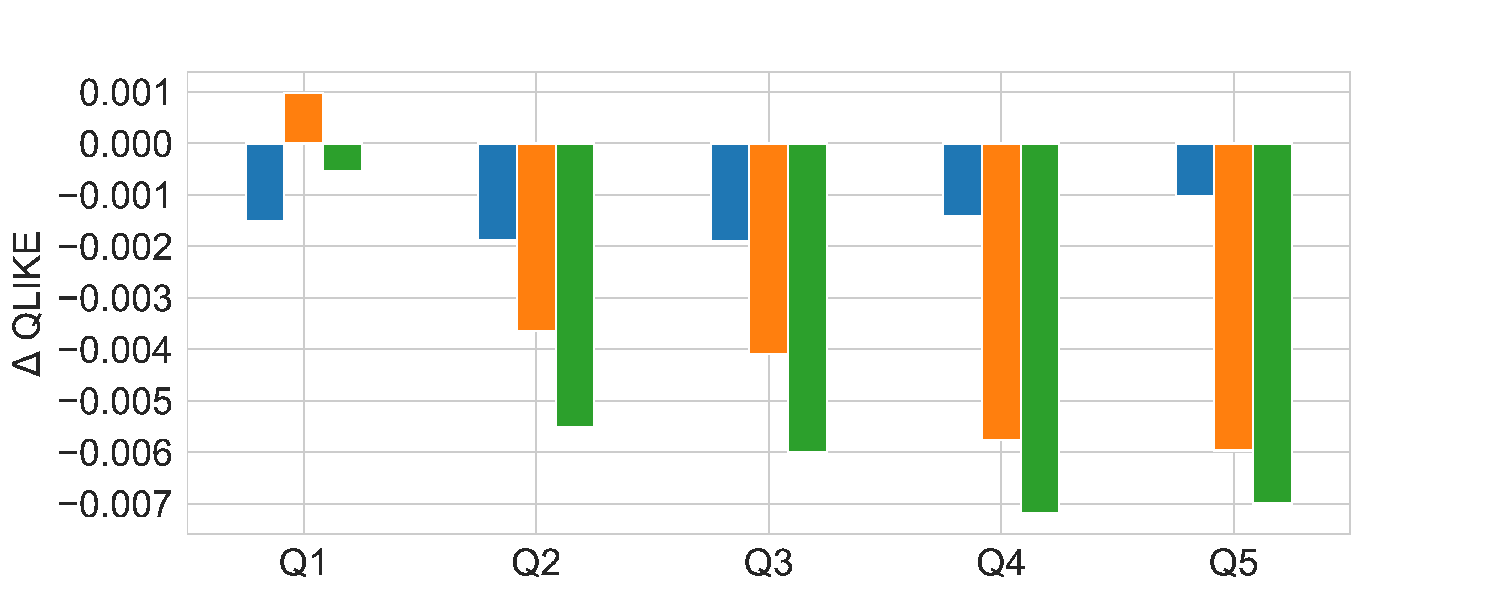}}
\subfigure[OLS, 1-day\label{fig:diff_common_1day}]{\includegraphics[width=.47\textwidth, trim=0mm 0mm 1cm 1cm, clip]{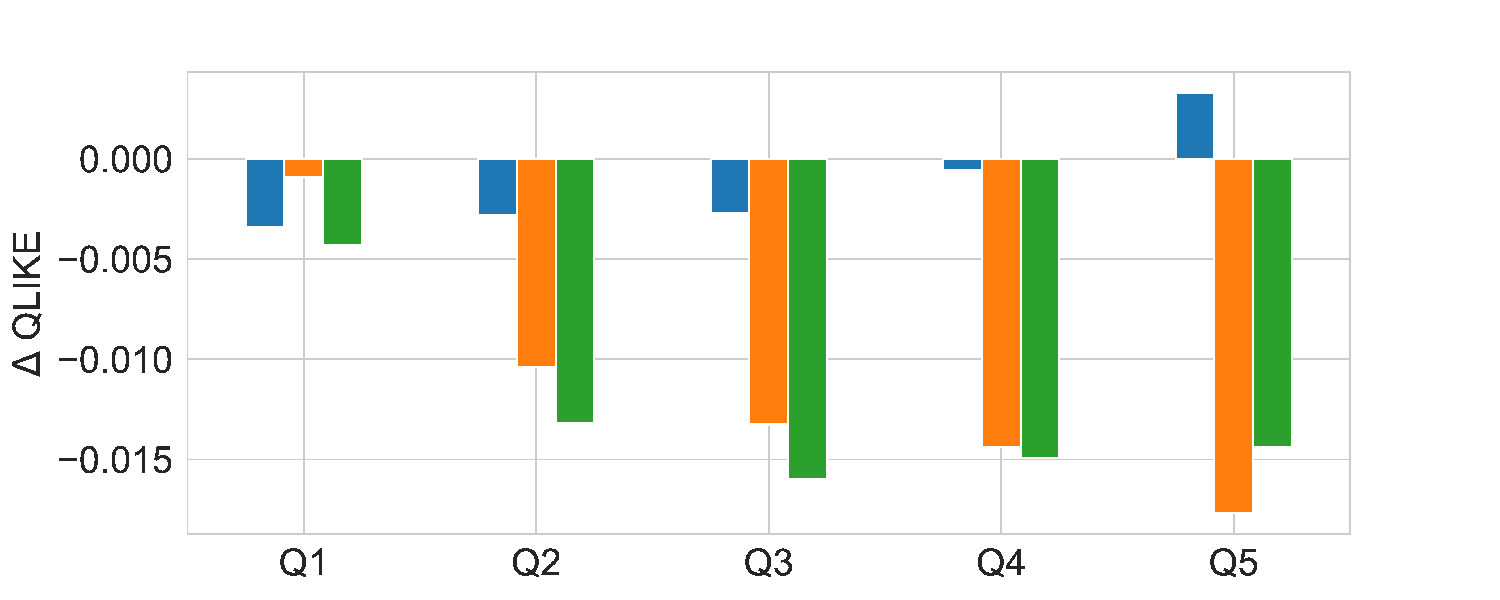}}
\caption{Pairwise  $\Delta$QLIKE of the OLS model sorted by commonality.} 
\caption*{\textit{Notes:} Q1, respectively Q5, denote the subset of stocks with the lowest, respectively highest, 20\% values for the commonality.}
\label{fig:diff_common}
\end{figure}

An interesting question to investigate is whether the improvements of {\univ} or {\aug} for individual stocks are associated with their commonality with the market volatility. To this end, we present the results in Figure \ref{fig:diff_common} for each quintile bucket, sorted by stock commonality (computed from Eqn \eqref{eqn:common_RV}). From this figure, we observe that the reduction of {\aug} in out-of-sample QLIKE relative to {\univ} is explained by commonality to a large extent. Generally, the out-of-sample QLIKE is expected to decline steadily for stocks with higher commonality. Another interesting result arises from Figure \ref{fig:diff_common_10min}, where we observe that {\univ} and {\aug} actually reduce the out-of-sample QLIKE more for stocks (in the Q1 bucket) that are most loosely connected to the market in terms of 10-min volatility. Further research should be undertaken to investigate this finding.

\subsection{Variable importance \& interaction effects}\label{sec:interaction}
This section provides intuition for why neural networks perform as strongly as they do, with an eye towards explainability. Due to the use of non-linear activation functions and multiple hidden layers,  neural networks enjoy the benefit of allowing for potentially complex interactions among predictors, albeit at the cost of considerably reducing model interpretability. To better understand such a ``black-box'' technique, we provide the following analysis to help illustrate the inner workings of neural networks and explain their competitive performance.

\paragraph{Relative importance of predictors.} In order to identify which variables are the most important for the prediction task at hand, we construct  a metric (see \citet{sadhwani2021deep}) based on the sum of absolute partial derivatives (Sensitivity) of the predicted volatility. In particular, to quantify the importance of the $k$-th predictor, we compute

\vspace{-3mm}
\begin{equation}\label{eq:sensitivity}
\mathrm{Sensitivity}_{k}= \sum_{i=1}^{N} \sum_{t \in \mathcal{T}_{train}} \bigg|\frac{\partial F}{\partial u_{k}}\bigg|_{\mathbf{u}=\mathbf{u}_{i,t}}
\end{equation}
\noindent Here, $F$ is the fitted model under the {\aug} scheme, $\mathbf{u}$ represents the vector of predictors and $u_{k}$ is the $k$-th element in $\mathbf{u}$. $\mathbf{u}_{i,t}$ represents the input features of stock $i$ at time $t$. We normalize the sensitivity of all variables such that they sum up to one. In a special case of linear regression, the sensitivity measure is the normalized absolute slope coefficient.

\begin{figure}[H]
\centering
\subfigure[OLS, 65-min\label{fig:ssd_ols_65min}]{\includegraphics[width=.48\textwidth, trim=2cm 0mm 3cm 5mm,clip]{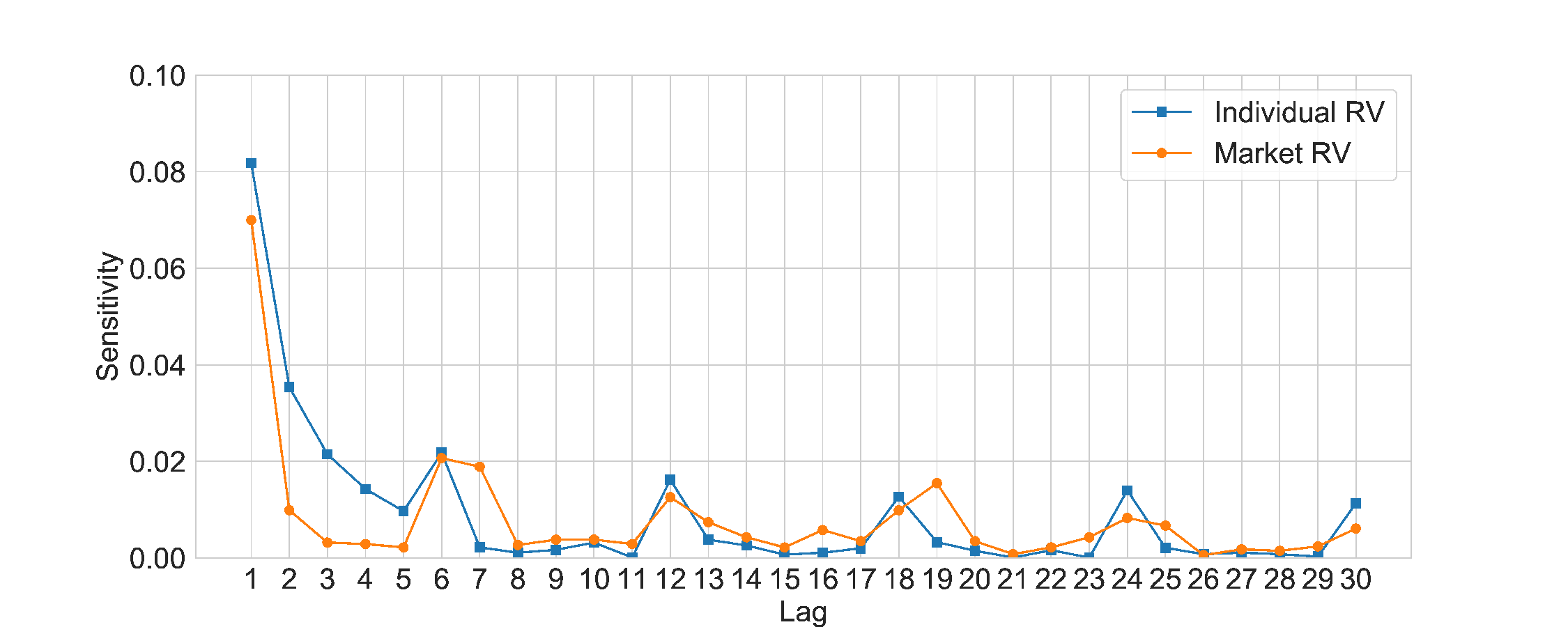}}
\subfigure[MLP, 65-min\label{fig:ssd_dnn_65min}]{\includegraphics[width=.48\textwidth, trim=2cm 0mm 3cm 5mm,clip]{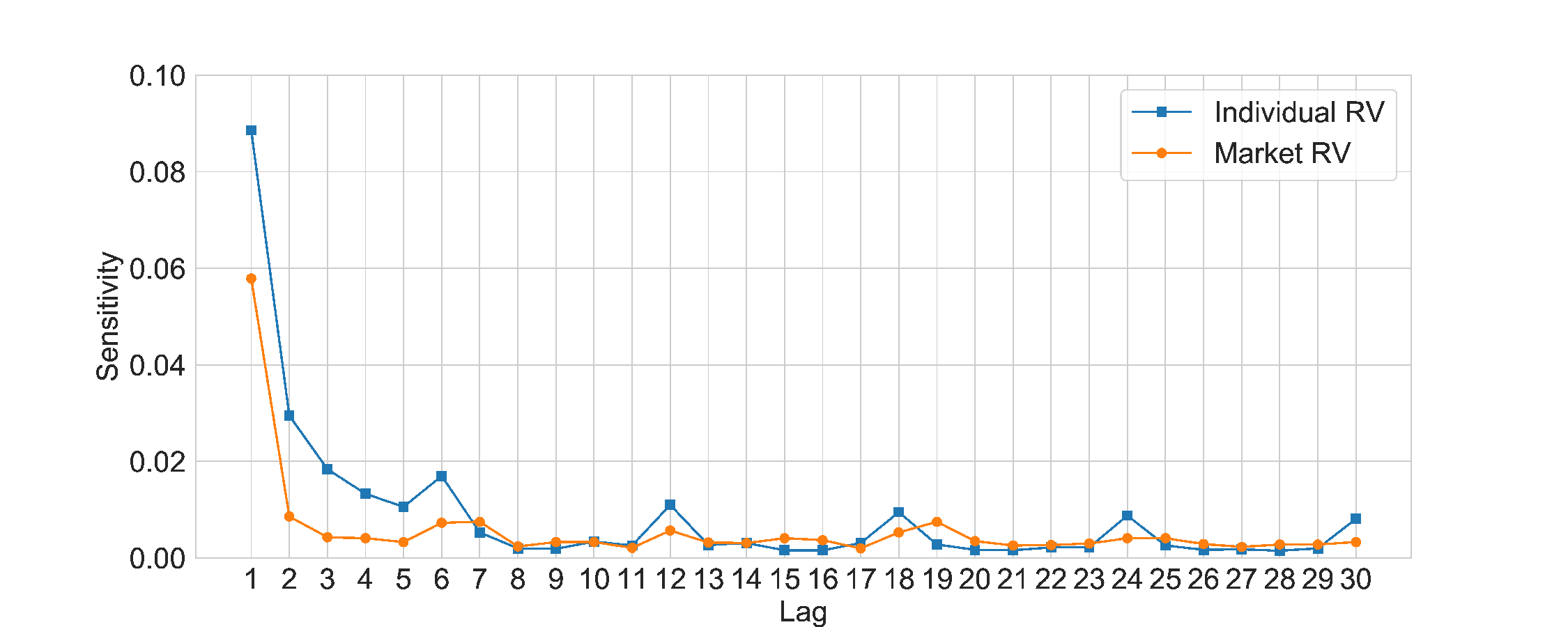}}
\caption{Relative importance of lagged individual and market RVs.}
\caption*{\textit{Notes:} For ease of readability, we only report the sensitivity values for the most recent 30 lagged RVs (i.e. in the last 5 days for 65-min horizon).}
\label{fig:sensitivity}
\end{figure}

Considering the 65-min scenario as an example, Figure \ref{fig:sensitivity} reveals that for both OLS and MLP, there has been a tendency of the lagged features to decline in terms of sensitivity, as the lag increases. Additionally, we observe that the sensitivity values rise to a high point at every 6 lags, corresponding to 1 day. A distinct difference between the sensitivity values implied by OLS and the ones implied by MLP is that the latter places more weight on the lag=1 individual RV (Sensitivity=0.90) and less on the lag=1 market RV (Sensitivity=0.059). On the other hand, for OLS, the sensitivities of lag=1 individual (resp. market) RV are 0.081 (resp. 0.069).

\paragraph{Interaction effects.} To analyze the interactions between the two most significant features implied by neural networks, we adopt an approach (e.g. \citet{gu2020empirical, choi2021alpha}) that focuses  on the partial relations between a pair of input variables and responses, while fixing other variables at their mean values

\vspace{-4mm}
\begin{equation}\label{eq:interaction}
{F}_{j \mid i}\left(u_{j} \mid u_{i}=q\right)=F\left(u_{j}, u_{i}=q, u_{k}=\bar{u}_{k}, k \neq i, j\right), 
\end{equation}

\noindent where $q$ represent the quantile values for the $i$-th predictor $u_{i}$.

\begin{figure}[H]
\centering
\subfigure[OLS, 65-min\label{fig:interact_ols_65min}]{\includegraphics[width=.48\textwidth, trim=5mm 0mm 1cm 5mm,clip]{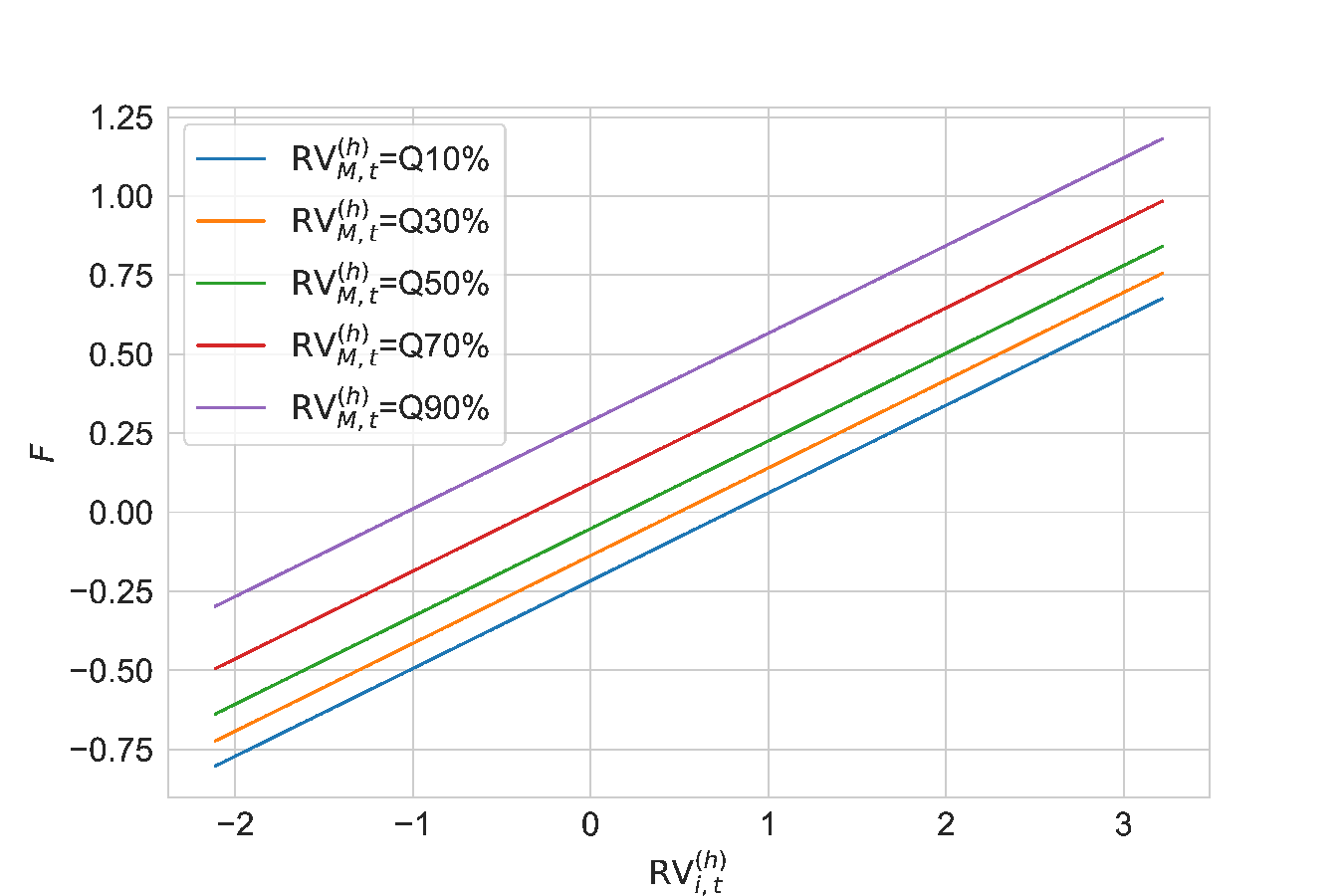}}
\subfigure[MLP, 65-min\label{fig:interact_dnn_65min}]{\includegraphics[width=.48\textwidth, trim=5mm 0mm 1cm 5mm,clip]{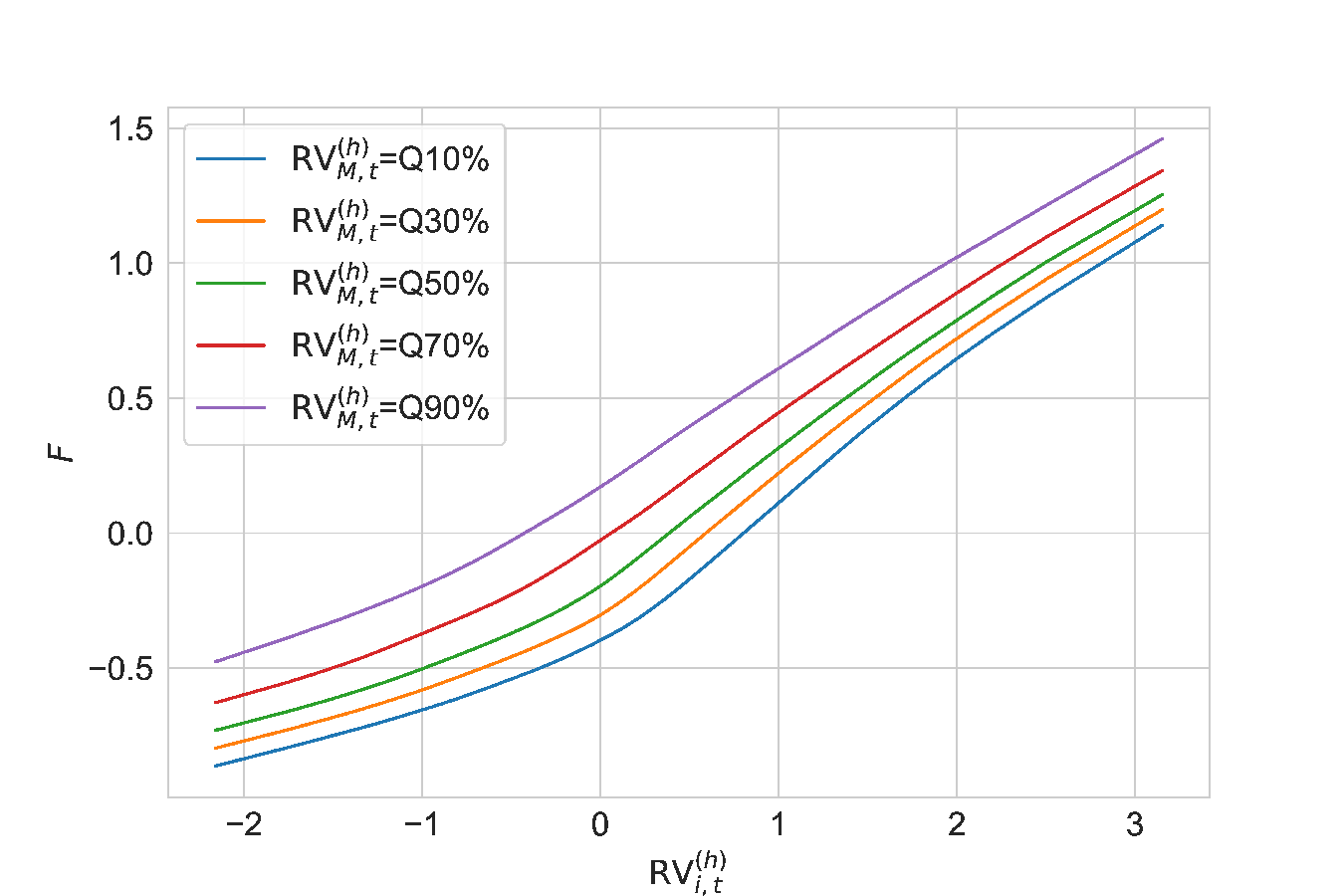}}
\caption{Interactions between the lagged individual and market RV.}
\caption*{\textit{Notes:} The figure plots the pattern of predicted RV ($y$-axis) as a function of the lag=1 individual RV ($x$-axis) conditioned on various lag=1 market RV quantile values (keeping all other variables at their mean values).}
\label{fig:interaction}
\end{figure}

Figure \ref{fig:interaction} illustrates how predicted volatility (i.e., the fitted values) varies as a function of the pair of predictor variables $\text{RV}_{i,t}^{(h)}$ and $\text{RV}_{M,t}^{(h)}$, over their support.\footnote{Recall that the variables are normalized by removing the mean and scaling to unit variance.} In particular, we analyze the interaction of the lag=1 individual RV ($\text{RV}_{i,t}^{(h)}$), with the lag=1 market RV ($\text{RV}_{M,t}^{(h)}$). As shown in Figure \ref{fig:interact_ols_65min}, a parallel shift between the different curves occurs if there are no interaction effects between  $\text{RV}_{i,t}^{(h)}$ and $\text{RV}_{M,t}^{(h)}$. Figure \ref{fig:interact_dnn_65min}  first reveals that the predicted volatility is non-linear in $\text{RV}_{i,t}^{(h)}$, and the slope of that relationship becomes higher after $\text{RV}_{i,t}^{(h)}$ exceeds a certain threshold (around 0.5). Furthermore, it demonstrates clear interaction effects between $\text{RV}_{i,t}^{(h)}$ and $\text{RV}_{M,t}^{(h)}$. As it can be observed from the rightmost region of Figure \ref{fig:interact_dnn_65min}, the distances between the curves become relatively smaller, conveying the message that, when an individual stock is very volatile, the market effect on it weakens.

\subsection{Forecasting RVs of unseen stocks}
To examine the ability to generalize and address concerns regarding  overfitting, we perform a stringent out-of-sample test, i.e. using the existent trained models to forecast the volatility of new stocks that have not been included in the training sample, in the spirit of \citet{sirignano2019universal,choi2021alpha}. For better distinction, we denote the stocks used for estimating machine learning models as \textbf{raw stocks}, and those new stocks not in the training sample as \textbf{unseen stocks}.\footnote{The set of unseen stocks includes the following 16 tickers: AMAT,  APD, BIIB, COF, DE, EQIX, EW, GPN, HUM, ICE, ILMN, ITW, NOC, NSC, PLD, SLB.} We follow the procedure of training, validation, and testing periods described in Section \ref{sec:setup}. Specifically, to predict the RVs of unseen stocks in a particular year, we train and validate the models using the past data of raw stocks exclusively.

In this experiment, we choose OLS models trained for each unseen stock as the baseline. The results are shown in Table \ref{tab:unseen_metaaug}. Note that models trained under {\single} cannot be applied to forecast unseen stocks, {since they are trained for each specific raw stock individually.} From Table \ref{tab:unseen_metaaug}, we conclude that NNs trained on the pooled data of raw stocks have better forecasting performance compared to baselines, across all horizons. This presents new empirical evidence for a \textit{universal volatility mechanism} among stocks. Furthermore, neural networks significantly outperform other methods across three metrics, over 10-min, 30-min, and 65-min forecasting horizons, thus validating their robustness.
Concerning the 1-day scenario, neural networks obtain comparable  results (QLIKE=0.252) to the best non-neural network model (MSE=0.249, attained by LASSO). The realized utility of the different risk models echoes that of the out-of-sample QLIKE.
  
\begin{table}[H]
\centering
\caption{Performance on unseen stocks.}
\resizebox{1.0\textwidth}{!}{\begin{threeparttable}   
\begin{tabular}{l l c c c c c c c c }
    \toprule
  \multicolumn{2}{c}{\textbf{ Panel A:} }  & \multicolumn{2}{c}{{10-min}} & \multicolumn{2}{c}{{30-min}} & \multicolumn{2}{c}{{65-min}} & \multicolumn{2}{c}{{1-day}} \\
     \cmidrule(lr){3-4}\cmidrule(lr){5-6}\cmidrule(lr){7-8}\cmidrule(lr){9-10}
     
 \multicolumn{2}{c}{\textbf{ Statistical performance} } & MSE & QLIKE & MSE & QLIKE & MSE & QLIKE & MSE & QLIKE \\ \midrule
\textcolor{aoe}{OLS}&\textcolor{aoe}{Unseen} & \textcolor{aoe}{0.664} & \textcolor{aoe}{0.372} &\textcolor{aoe}{0.329}  &\textcolor{aoe}{0.219}  & \textcolor{aoe}{0.287} & \textcolor{aoe}{0.205} & \textcolor{aoe}{0.348} & \textcolor{aoe}{0.254}\\ \midrule
 & {\univ}                     & 0.678 & 0.410 &0.328  &0.223  & 0.286 & 0.206 & 0.343 & 0.260 \\
\multirow{-2}{*}{OLS} & {\aug} & 0.639 & 0.359 & 0.317 &0.222  & 0.278 & 0.208 & \textcolor{red}{0.327}* & \textcolor{red}{0.249}* \\ \midrule
 
 & {\univ}                       & 0.683 & 0.419 &0.330  &0.225  & 0.286 & 0.207 & 0.344 & 0.261 \\
\multirow{-2}{*}{LASSO} & {\aug} & 0.639 & 0.359 & 0.317 &0.222  & 0.278 & 0.208 & \textcolor{blue}{0.327}* & \textcolor{blue}{0.249}* \\ \midrule
 & {\univ}                       & 0.655 & 0.476  &0.314  & 0.206 &0.278 & 0.201 & 0.353 & 0.266 \\
\multirow{-2}{*}{XGBoost} & {\aug} & 0.654 & 0.509  &0.320  & 0.221 &0.282 & 0.206 & 0.364 & {0.255} \\ \midrule

 & {\univ}                       & \textcolor{blue}{0.623} & \textcolor{blue}{0.328}  &\textcolor{blue}{0.306}  & \textcolor{blue}{0.203}* &\textcolor{red}{0.266} & \textcolor{red}{0.193}* & 0.342 & 0.266 \\
\multirow{-2}{*}{MLP} & {\aug}   & 0.623 & 0.332  &\textcolor{red}{0.301}*  & \textcolor{red}{0.203}* &\textcolor{blue}{0.263}* & \textcolor{blue}{0.194}* & 0.329 & 0.252 \\ \midrule
  & {\univ}                       & 0.637 & 0.348  &0.311  & 0.211 &0.267 & 0.195 & 0.339 & 0.265 \\
\multirow{-2}{*}{LSTM} & {\aug}   & \textcolor{red}{0.622}* & \textcolor{red}{0.326}*  &0.303  & 0.205 & 0.263* & 0.194 & 0.332 & 0.255 \\ 
\midrule
 \multicolumn{2}{c}{\textbf{ Panel B:} }  & \multicolumn{2}{c}{{10-min}} & \multicolumn{2}{c}{{30-min}} & \multicolumn{2}{c}{{65-min}} & \multicolumn{2}{c}{{1-day}} \\
     \cmidrule(lr){3-4}\cmidrule(lr){5-6}\cmidrule(lr){7-8}\cmidrule(lr){9-10}
     
 \multicolumn{2}{c}{\textbf{ Realized utility} } & RU & RU-TC & RU & RU-TC & RU & RU-TC & RU & RU-TC \\ \midrule
 \textcolor{aoe}{OLS}&\textcolor{aoe}{Unseen} & \textcolor{aoe}{3.107} & \textcolor{aoe}{1.996} &\textcolor{aoe}{3.475}  &\textcolor{aoe}{2.672}  & \textcolor{aoe}{3.503} & \textcolor{aoe}{2.715} & \textcolor{aoe}{3.385} & \textcolor{aoe}{3.320}\\ \midrule
 & {\univ}                     & 2.988 & 2.280 &3.461  &2.700  & 3.498 & 2.712 & 3.363 & 3.298 \\
\multirow{-2}{*}{OLS} & {\aug} & 3.138 & 2.355 &3.459  &2.710  & 3.487 & 2.712 & \textcolor{blue}{3.389} & \textcolor{blue}{3.311} \\ \midrule
 
 & {\univ}                       & 2.959 & 2.270 &3.457  &2.704  & 3.496 & 2.712 & 3.359 & 3.296 \\
\multirow{-2}{*}{LASSO} & {\aug} & 3.137 & 2.376 & 3.458 &2.720  & 3.485 & 2.716 & \textcolor{red}{3.389} & \textcolor{red}{3.315} \\ \midrule
 & {\univ}                       & 2.688 & 1.640  &3.510  &2.711 &3.511 & 2.701 & 3.349 & 3.269 \\
\multirow{-2}{*}{XGBoost} & {\aug} & 2.563 &1.578  &3.464  &2.688  &3.495 & 2.680 &3.388 & 3.302 \\ \midrule

 & {\univ}                       & \textcolor{blue}{3.233} & \textcolor{blue}{2.396}  &\textcolor{red}{3.515}  & \textcolor{red}{2.736} &\textcolor{red}{3.529} & \textcolor{red}{2.730} & 3.340 & 3.266 \\
\multirow{-2}{*}{MLP} & {\aug}   & 3.221 & 2.444  &\textcolor{blue}{3.514}  & \textcolor{blue}{2.749} &3.522 & 2.735 & 3.378 & 3.302 \\ \midrule
  & {\univ}                       & 3.167 & 2.415  &3.493  & 2.769 &3.523 & 2.737 & 3.345 & 3.271 \\
\multirow{-2}{*}{LSTM} & {\aug}   & \textcolor{red}{3.238} & \textcolor{red}{2.533}  &3.507  & 2.787 &\textcolor{blue}{3.524} & \textcolor{blue}{2.762} & 3.371 & 3.302 \\

\bottomrule

\end{tabular}
\end{threeparttable}}
\caption*{\textit{Notes:} The table reports the out-of-sample results for predicting future realized volatility of unseen stocks over multiple horizons using different models under three training schemes. The row \textcolor{aoe}{OLS Unseen} represents the baseline results based on OLS models estimated for each unseen stock. Other rows represent the results of models estimated on raw stocks under the {\univ} and  {\aug} settings. For each horizon, the model with the best (second best) out-of-sample performance in terms of QLIKE (in Panel A) / RU (in Panel B) is highlighted in \textcolor{red}{red} (\textcolor{blue}{blue}), respectively. An asterisk (*) indicates models that are included in the MCS at the 5\% significance level.}
\label{tab:unseen_metaaug}
\end{table}

\section{Forecasting daily RVs with intraday RVs}\label{sec:intra2daily}
Given the fact that intraday volatility exhibits a high and  stable commonality (see Sections \ref{sec:data} and \ref{sec:commonality}), we are interested in the potential  benefits of using past intraday RVs to forecast daily RVs.

\subsection{Closely related literature} 
Generally speaking, there are two broad families of models used to forecast daily volatility: (i) GARCH and stochastic volatility (SV) models that employ daily returns; and (ii) models that use daily RVs. 
{Previous well-established studies have} shown that due to the utilization of available intraday information, daily realized volatility is a superior proxy for the unobserved daily volatility, when compared to the parametric volatility measures  generated from the GARCH and SV models of daily returns (see \citet{andersen2003modeling, barndorff2002econometric, izzeldin2019forecasting}). It is worth noting that in these traditional forecasting daily RV models (e.g. ARFIMA of \citet{andersen2003modeling}, HAR of \citet{corsi2009simple}, SHAR of \citet{patton2015good}, HARQ of \citet{bollerslev2016exploiting}), only past daily RVs (or their alternatives) are included as predictors. {Even though this is a mainstream approach in the literature, it does not  benefit to the full extent from the availability of intraday data.} In the presidential address of SoFiE 2021, \citet{bollerslev2022realized} also pointed out that ``semivariation measured over shorter interday time intervals may afford additional useful information.'' 

{Intraday RV information is also studied for forecasting the one-day-ahead volatility in several previous works, e.g. the mixed data sampling (MIDAS) approach of \citet{ghysels2005there, ghysels2006predicting} and the ``Rolling'' approach of \citet{pascalau2021increasing}. In particular, the classic MIDAS approach uses smooth-distributed lag polynomials of high-frequency predictors to forecast the low-frequency target variables, in the form $\text{RV}_{i,t+1}^{(d)}=\beta_{i,0}+\beta_{i,1}\left[a(1)^{-1} a(L)\right] \text{RV}_{i,t}^{(d)}+\epsilon_{i,t+1}$, where the $a(L)$ lag polynomial is defined by scaled beta functions. \citet{ghysels2006predicting} find that the direct use of high-frequency data does not improve volatility predictions compared to the forecasts from a model based on daily RVs only. We reckon it is due to the restricted flexibility of MIDAS models, usually with one or two parameters determining the pattern of the weights, therefore missing the time-of-day effect of intraday RVs.  \citet{pascalau2021increasing} increase the training samples by rolling a fixed window of intraday returns over consecutive trading days by adding and dropping one intraday return at each end. They claim that their proposed ``Rolling'' approach could potentially capture the changing dynamics of serial correlation throughout the trading day, thus leading to improved volatility forecasts.}

\subsection{Proposed approach} 
{In Section \ref{sec:models}, we introduced a set of commonly used models, where the daily variables (lagged daily RVs) are employed as predictors when forecasting 1-day RVs. For simplicity, we refer to these models as \textbf{traditional} approaches in this section. Previous sections, such as Figure \ref{fig:sensitivity}, concluded that the most recent RV plays a more important role in forecasting future volatility. Motivated by the fact that intraday volatility has a high and stable commonality, we propose a new prediction approach for forecasting daily volatility  using past intraday RVs as predictors, denoted by \textbf{Intraday2Daily} approach.}

\begin{figure}[H]
\centering
\includegraphics[width=.71\textwidth, trim=3.5cm 4cm 3.5cm 3cm,clip]{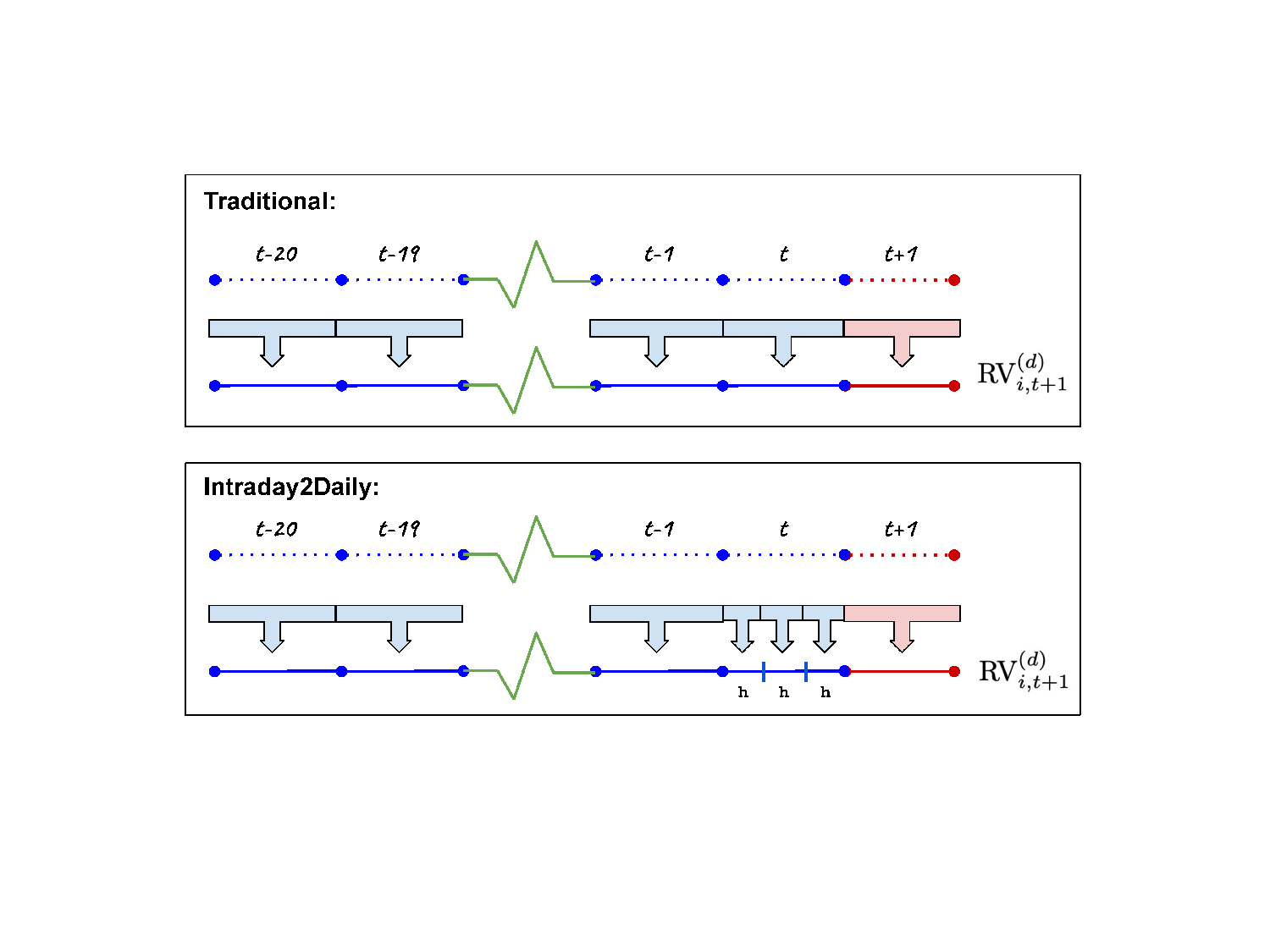}
\caption{Illustration of two prediction approaches for future daily volatility (\textcolor{red}{red} segment).}
\caption*{\textit{Notes:} In each box, dots in the top line represent the intraday returns. The traditional approaches employ the aggregated daily (or weekly, or monthly) RVs (\textcolor{blue}{long blue} segments) as predictors, while the \textbf{Intraday2Daily} approach employs intraday RVs (\textcolor{blue}{short blue} segments between two adjacent vertical ticks). $h$ represents the horizon of intraday RVs. In this example, $h=130$ minutes.}
\label{fig:intra2daily}
\end{figure}

In contrast to \citet{ghysels2006predicting, pascalau2021increasing}, our \textbf{Intraday2Daily} approach takes the time-of-day effect into account in an explicit way, {and posits the model} 
\begin{equation}\label{eq:intra}
    \text{RV}_{i,t+1}^{(d)} = {F}_i\left(\text{RV}_{i,t}^{(h)}, \dots, \text{RV}_{i,t-(p-1)h}^{(h)}, \text{RV}_{i, t-1}^{(d)}, \dots, \text{RV}_{i, t-(p-1)}^{(d)}; \theta\right) + \epsilon_{i,t+1}.
\end{equation}
Here $\left(\text{RV}_{i,t}^{(h)}, \dots, \text{RV}_{i,t-(k-1)h}^{(h)}\right)$ represent the past RVs for stock $i$ computed over shorter intraday time horizons $h$ at day $t$ and $\left(\text{RV}_{i, t-1}^{(d)}, \dots, \text{RV}_{i, t-(p-1)}^{(d)}\right)$ are past daily RVs of stock $i$ up to day $t-1$.
Departing from traditional models where all the variables are computed in the daily frequency, we decompose the lag-one daily $\text{RV}_{i, t}^{(d)}$ to sub-sampled RVs, i.e. $\left(\text{RV}_{i,t}^{(h)}, \dots, \text{RV}_{i,t-(k-1)h}^{(h)}\right)$. Under the {\aug} training scheme, we also incorporate the market volatilities into models. Figure \ref{fig:intra2daily} illustrates the comparison between the traditional approach and our \textbf{Intraday2Daily} approach.

The advantages of the \textbf{Intraday2Daily} approach over  traditional approaches can be summarized as follows. First, the \textbf{Intraday2Daily} approach significantly enriches the information content of daily volatility. Second, it contributes to the literature in the modeling of daily volatility by examining the coefficients of intraday RVs.
Third, the essential idea underlying the \textbf{Intraday2Daily} approach can be possibly applied to estimate other daily risk measures, such as value-at-risk (VaR), etc. For example, one may use half-hour VaRs to forecast the one-day-ahead VaR.
Finally, practitioners can better adjust their portfolios with more accurate forecasts from the \textbf{Intraday2Daily} approach rather than traditional approaches.
To the best of our knowledge, this is the first study to explicitly investigate the predictive power of intraday RVs on daily volatility, and to demonstrate the additional accuracy improvements it brings to the forecasting task.

\subsection{Experiments}
The forecasting performance of traditional approaches with daily variables are already summarized in the column '1-day' of Table \ref{tab:experiment_sigmetaaug}. Table \ref{tab:experiment_intra2daily} reports the results of models combined with the \textbf{Intraday2Daily} approach.\footnote{We observe similar findings when applying the \textbf{Intraday2Daily} approach to forecast the raw volatilities (not in logs).} In other words, models in Table \ref{tab:experiment_intra2daily} use sub-sampled intraday RVs rather than the lag-one total RV in the column '1-day' of Table \ref{tab:experiment_sigmetaaug}. For example, the lag-one total RV in HAR (Eqn \eqref{eq:har}) is replaced by non-overlapped intraday RVs. 

By comparing the column '1-day' of Table \ref{tab:experiment_sigmetaaug} with Table  \ref{tab:experiment_intra2daily}, we establish that the \textbf{Intraday2Daily} approach generally helps improve the out-of-sample performance of benchmark models. For example, under the {\single} setting, compared to OLS using daily RVs (QLIKE=0.192), 65-min RVs improve the out-of-sample performance (QLIKE=0.186). 

MLPs with intraday RVs again achieve the best out-of-sample performance. For example, the QLIKEs of MLPs under {\univ} are $\{0.171, 0.171, 0.172\}$ using $\{10\mbox{-min}, 30\mbox{-min}, 65\mbox{-min}\}$ RVs as predictors, respectively. The superior performance of MLPs over linear regressions when using intraday RVs further demonstrates the advantages of NNs to learn unknown dynamics in financial markets. 

In general, the improvements of the \textbf{Intraday2Daily} approach lead to higher utilities. For instance, when considering the column '1-day' of Panel B in Table \ref{tab:experiment_sigmetaaug}, we observe that the RU (resp. RU-TC) values of OLS under {\aug} are 3.576\% (resp. 3.536\%). OLS with 65-min RVs as predictors obtain higher RU (resp. RU-TC) values of 3.583\% (resp. 3.547\%). Overall, MLPs deliver the highest utility (RU=3.593\%, RU-TC=3.550\%) based on the 65-min intraday RVs, followed by LSTMs, thus hinting at potential non-linearity and complex interactions inherent in the data.

\begin{table}[]
\centering
\caption{Out-of-sample performance of the \textbf{Intraday2Daily} approach.}
\resizebox{0.8\textwidth}{!}{\begin{threeparttable}   
\begin{tabular}{l l c c c c c c c c }
    \toprule
  \multicolumn{2}{c}{\textbf{ Panel A:} }  & \multicolumn{2}{c}{{10-min}} & \multicolumn{2}{c}{{30-min}} & \multicolumn{2}{c}{{65-min}} \\
     \cmidrule(lr){3-4}\cmidrule(lr){5-6}\cmidrule(lr){7-8}
     
 \multicolumn{2}{c}{\textbf{ Statistical performance} } 
  & MSE & QLIKE & MSE & QLIKE & MSE & QLIKE \\ \midrule
 & {\single} & 0.259 & 0.189 & 0.252 & 0.185 & 0.252 & 0.185 \\
 & {\univ} & 0.270 & 0.197 & 0.255 & 0.187 & 0.253 & 0.186 \\
\multirow{-3}{*}{HAR} & {\aug} & 0.256 & 0.179 & 0.249 & 0.174 & 0.249 & 0.175 \\\midrule
 & {\single} & 0.255 & 0.186 & 0.252 & 0.186 & 0.253 & 0.186 \\
 & {\univ} & 0.253 & 0.186 & 0.252 & 0.187 & 0.252 & 0.186 \\
\multirow{-3}{*}{OLS} & {\aug} & 0.249 & 0.173 & 0.248 & 0.173 & 0.249 & 0.173 \\\midrule
 & {\single} & 0.262 & 0.194 & 0.251 & 0.185 & 0.253 & 0.186 \\
 & {\univ} & 0.261 & 0.191 & 0.248 & 0.187 & 0.248 & 0.186 \\
\multirow{-3}{*}{LASSO} & {\aug} & 0.273 & 0.203 & 0.247 & 0.173 & 0.247 & 0.173 \\\midrule
 & {\single} & 0.323 & 0.204 & 0.330 & 0.201 & 0.332 & 0.200 \\
 & {\univ} & 0.261 & 0.177 & 0.257 & 0.173 & 0.257 & 0.173 \\
\multirow{-3}{*}{XGBoost} & {\aug} & 0.264 & 0.179 & 0.261 & 0.176 & 0.266 & 0.176 \\\midrule
 & {\single} & - & - & - & - & - & - \\
 & {\univ} & \textcolor{blue}{0.243}* & \textcolor{blue}{0.171}* & \textcolor{red}{0.242}* & \textcolor{red}{0.171}* & 0.246 & 0.172 \\
\multirow{-3}{*}{MLP} & {\aug} & 0.247 & 0.174 & 0.246 & 0.175 & 0.247 & 0.176 \\\midrule
 & {\single} & - & - & - & - & - & - \\
 & {\univ} & 0.247 & 0.174 & 0.244 & 0.171* & 0.244 & 0.171 \\
\multirow{-3}{*}{LSTM} & {\aug} & 0.258 & 0.184 & 0.249 & 0.175 & 0.250 & 0.176 \\\midrule

  \multicolumn{2}{c}{\textbf{ Panel B:} }  & \multicolumn{2}{c}{{10-min}} & \multicolumn{2}{c}{{30-min}} & \multicolumn{2}{c}{{65-min}} \\
     \cmidrule(lr){3-4}\cmidrule(lr){5-6}\cmidrule(lr){7-8}
     
 \multicolumn{2}{c}{\textbf{ Realized utility} } 
 &RU & RU-TC & RU & RU-TC & RU & RU-TC\\ \midrule
 & {\single} & 3.558 & 3.524 & 3.567 & 3.533 & 3.566 & 3.531 \\
 & {\univ} & 3.539 & 3.521 & 3.563 & 3.534 & 3.565 & 3.532 \\
\multirow{-3}{*}{HAR} & {\aug} & 3.562 & 3.532 & 3.573 & 3.541 & 3.570 & 3.536 \\\midrule
 & {\single} & 3.565 & 3.526 & 3.565 & 3.530 & 3.564 & 3.529 \\
 & {\univ} & 3.565 & 3.534 & 3.564 & 3.535 & 3.565 & 3.533 \\
\multirow{-3}{*}{OLS} & {\aug} & 3.581 & 3.548 & 3.583 & 3.551 & 3.583 & 3.547 \\\midrule
 & {\single} & 3.561 & 3.526 & 3.565 & 3.531 & 3.564 & 3.529 \\
 & {\univ} & 3.561 & 3.524 & 3.564 & 3.535 & 3.573 & 3.536 \\
\multirow{-3}{*}{LASSO} & {\aug} & 3.551 & 3.515 & 3.565 & 3.531 & 3.585 & 3.548 \\\midrule
 & {\single} & 3.532 & 3.474 & 3.545 & 3.486 & 3.550 & 3.489 \\
 & {\univ} & 3.584 & 3.532 & 3.593 & 3.543 & {3.590} & {3.547} \\
\multirow{-3}{*}{XGBoost} & {\aug} & 3.579 & 3.527 & 3.586 & 3.535 & 3.589 & 3.538 \\\midrule
 & {\single} & - & - & - & - & - & - \\
 & {\univ} & 3.586 & 3.543 & 3.592 & 3.549 & \textcolor{red}{3.593} & \textcolor{red}{3.550} \\
\multirow{-3}{*}{MLP} & {\aug} & 3.562 & 3.521 & 3.583 & 3.541 & 3.581 & 3.540 \\\midrule
 & {\single} & - & - & - & - & - & - \\
 & {\univ} & 3.586 & 3.543 & 3.592 & 3.549 & \textcolor{blue}{3.593} & \textcolor{blue}{3.550} \\
\multirow{-3}{*}{LSTM} & {\aug} & 3.562 & 3.521 & 3.583 & 3.541 & 3.581 & 3.540 \\
\bottomrule
\end{tabular}
\end{threeparttable}    }
\label{tab:experiment_intra2daily}
\caption*{\textit{Notes:} The table reports the out-of-sample results for predicting future daily realized volatility using different models under three training schemes when combined with the \textbf{Intraday2Daily} approach. The columns ('10-min', '30-min', and '65-min') represent the frequency of predictor features and the dependent variable in this table always corresponds to future daily volatility. The model with the best (second best) out-of-sample performance in QLIKE (in Panel A) / RU (in Panel B) is highlighted in \textcolor{red}{red} (\textcolor{blue}{blue}), respectively. An asterisk (*) indicates models that are included in the MCS at the 5\% significance level.}
\end{table}

\subsection{Robustness check}
In this section, we present the empirical analysis of examining the robustness of the  \textbf{Intraday2Daily} approach when incorporating new types of predictors (including semi-RV of \citet{patton2015good}, and realized quarticity of \citet{bollerslev2016exploiting}).

\paragraph{SHAR.} \citet{patton2015good} proposed the Semi-variance-HAR (SHAR) model as an extension of the standard HAR model (see further details in \ref{subsec:HAR}),  in order to exploit the well-documented leverage effect by decomposing the total RV of the first lag via signed intraday returns, as shown in Eqn \eqref{eq:shar_semi} (see \citet{barndorff2008measuring}). In other words, the lag-one RV in SHAR (Eqn \eqref{eq:shar}) is split into the sum of squared positive returns and the sum of squared negative returns, as follows. 

\begin{equation}\label{eq:shar_semi}
\begin{array}{l}
\mathrm{RV}_{i,t}^{(d)+}=\sum_{l=0}^{M-1} r_{i,t-l \cdot \Delta}^{2} I_{\left\{r_{t-l \cdot \Delta}>0\right\}}, \\
\mathrm{RV}_{i,t}^{(d)-}=\sum_{l=0}^{M-1} r_{i, t-l \cdot \Delta}^{2} I_{\left\{r_{t-l \cdot \Delta}<0\right\}},
\end{array}
\end{equation}
\vspace{-5mm}
\begin{equation}\label{eq:shar}
\mathrm{RV}_{i, t+1}^{(d)} = \alpha_i + \beta_i^{(d)+} {\mathrm{RV}}_{i, t}^{(d)+} + \beta_i^{(d)-} {\mathrm{RV}}_{i, t}^{(d)-} + \beta_i^{(w)} {\mathrm{RV}}_{i, w}^{(w)}+ \beta_i^{(m)} {\mathrm{RV}}_{i, m}^{(m)}+\epsilon_{i, t+1}.
\end{equation}
\noindent In the above, $\Delta$ denotes the interval for computing the intraday returns.

\paragraph{HARQ.} 
\citet{bollerslev2016exploiting} pointed out that the beta coefficients in the HAR model may be affected by measurement errors in the realized volatilities. 
By exploiting the asymptotic theory for high-frequency realized volatility estimation, the authors propose an easy-to-implement model, termed as HARQ (Eqn \eqref{eq:harq}). The realized quarticity (RQ) is estimated according to Eqn \eqref{eq:quarticity}, aiming to correct the measurement errors. 

\begin{equation}\label{eq:quarticity}
\mathrm{RQ}_{i, t}^{(d)}=\frac{M}{3} \sum_{l=0}^{M-1} r_{t-l \cdot \Delta}^{4}
\end{equation}
\vspace{-5mm}
\begin{equation}\label{eq:harq}
\mathrm{RV}_{i, t+1}^{(d)} = \alpha_i +\left(\beta_i^{(d)}+\beta_i^{(d) \mathrm{Q}} \sqrt{\mathrm{RQ}_{i,t}^{(d)}}\right) \mathrm{RV}_{i,t}^{(d)}+\beta_i^{(w)} {\mathrm{RV}}_{i,t}^{(w)}+\beta_i^{(m)} {\mathrm{RV}}_{i,t}^{(m)}+\epsilon_{i, t+1}.
\end{equation}

We compute the corresponding intraday variables of semi-RVs and RQs and then include them as new predictors in the \textbf{Intraday2Daily} approach. From Table \ref{tab:experiment_intra2daily_SHAR}, we first observe that the SHAR model generally performs as well as the standard HAR model (in Table \ref{tab:experiment_sigmetaaug}), in line with \citet{bollerslev2016exploiting}. HARQ outperforms HAR and SHAR, when applied to individual stocks studied in the present paper. {Comparing the 'Traditional' column with others, we conclude that in general, replacing the daily RVs with intraday RVs as predictors helps improve the out-of-sample performance of benchmark models.}

\begin{table}[H]
\centering
\caption{Out-of-sample performance of the \textbf{Intraday2Daily} approach.}
\resizebox{1.0\textwidth}{!}{\begin{tabular}{llcccccccc }
    \toprule
  \multicolumn{2}{c}{\textbf{Panel A:} }  & \multicolumn{2}{c}{{10-min}} & \multicolumn{2}{c}{{30-min}} & \multicolumn{2}{c}{{65-min}} & \multicolumn{2}{c}{{Traditional}} \\
     \cmidrule(lr){3-4}\cmidrule(lr){5-6}\cmidrule(lr){7-8} \cmidrule(lr){9-10}
     
 \multicolumn{2}{c}{\textbf{ Statistical performance} } 
  & MSE & QLIKE & MSE & QLIKE & MSE & QLIKE & MSE & QLIKE \\ \midrule
 & {\single} & 0.277 & 0.191 & 0.257 & 0.178 & 0.253 & 0.176 & 0.261 & 0.183 \\
 & {\univ} & 0.285 & 0.198 & 0.263 & 0.183 & 0.255 & 0.178 & 0.261 & 0.182 \\
\multirow{-3}{*}{SHAR} & {\aug} & 0.261 & 0.181 & 0.253 & 0.175 & 0.250 & 0.174 & 0.254 & 0.178\\\midrule
 & {\single} & 0.264 & 0.204 & 0.254 & 0.178 & 0.253 & 0.176 & 0.256 & 0.179 \\
 & {\univ} & 0.253 & 0.176 & 0.253 & 0.176 & 0.254 & 0.176 & 0.257  & 0.179 \\
\multirow{-3}{*}{HARQ} & {\aug} & 0.251 & 0.174 & \textcolor{red}{0.248}* & \textcolor{red}{0.172}* & \textcolor{blue}{0.250} & \textcolor{blue}{0.174} & 0.253 & 0.176 \\\midrule

  \multicolumn{2}{c}{\textbf{Panel B:} }  & \multicolumn{2}{c}{{10-min}} & \multicolumn{2}{c}{{30-min}} & \multicolumn{2}{c}{{65-min}} & \multicolumn{2}{c}{{Traditional}} \\
     \cmidrule(lr){3-4}\cmidrule(lr){5-6}\cmidrule(lr){7-8} \cmidrule(lr){9-10}
 \multicolumn{2}{c}{\textbf{ Realized utility} } 
 &RU & RU-TC & RU & RU-TC & RU & RU-TC & RU & RU-TC \\ \midrule 

 & {\single} & 3.528 & 3.497 & 3.559 & 3.525 & 3.563 & 3.529 & 3.548 & 3.515 \\
 & {\univ} & 3.510 & 3.499 & 3.548 & 3.525 & 3.560 & 3.529 & 3.550 & 3.516 \\
\multirow{-3}{*}{SHAR} & {\aug} & 3.563 & 3.533 & 3.576 & 3.545 & 3.578 & 3.545 & 3.571 & 3.537 \\\midrule
 & {\single} & 3.467 & 3.425 & 3.556 & 3.520 & 3.564 & 3.528 & 3.557 & 3.525 \\
 & {\univ} & 3.564 & 3.530 & 3.564 & 3.530 & 3.564 & 3.530 & 3.558 & 3.525 \\
\multirow{-3}{*}{HARQ} & {\aug} & \textcolor{blue}{3.580} & \textcolor{blue}{3.544} & \textcolor{red}{3.583} & \textcolor{red}{3.546} & 3.578 & 3.541 & 3.575 & 3.538 \\
\bottomrule
\end{tabular}}
\label{tab:experiment_intra2daily_SHAR}
\caption*{\textit{Notes:} The table reports the out-of-sample results of SHAR and HARQ for predicting future daily realized volatility under three training schemes. Columns '10-min', '30-min', and '65-min' represent the \textbf{Intraday2Daily} approach with different frequencies of predictors while the column 'Traditional' represents that lagged daily RVs are used as predictors. The dependent variable in this table always corresponds to future daily volatility. The model with the best (second best) out-of-sample performance in QLIKE (in Panel A) / RU (in Panel B) is highlighted in \textcolor{red}{red} (\textcolor{blue}{blue}), respectively. An asterisk (*) indicates models that are included in the MCS at the 5\% significance level.}
\end{table}

\subsection{Analysis of the time-of-day dependent RV} 
To offer a more comprehensive understanding of the performance of 
\textit{time-of-day dependent} RVs, we examine the coefficients of the \textbf{Intraday2Daily} OLS model trained under {\aug}. Recall that before we input features into the model, we rescale them to have a mean of zero and a standard deviation of one. Hence we can compare the coefficients of different lagged variables.

\begin{figure}[H]
\centering
\includegraphics[width=0.9\textwidth, trim=4.5cm 5mm 5cm 2.5cm,clip ]{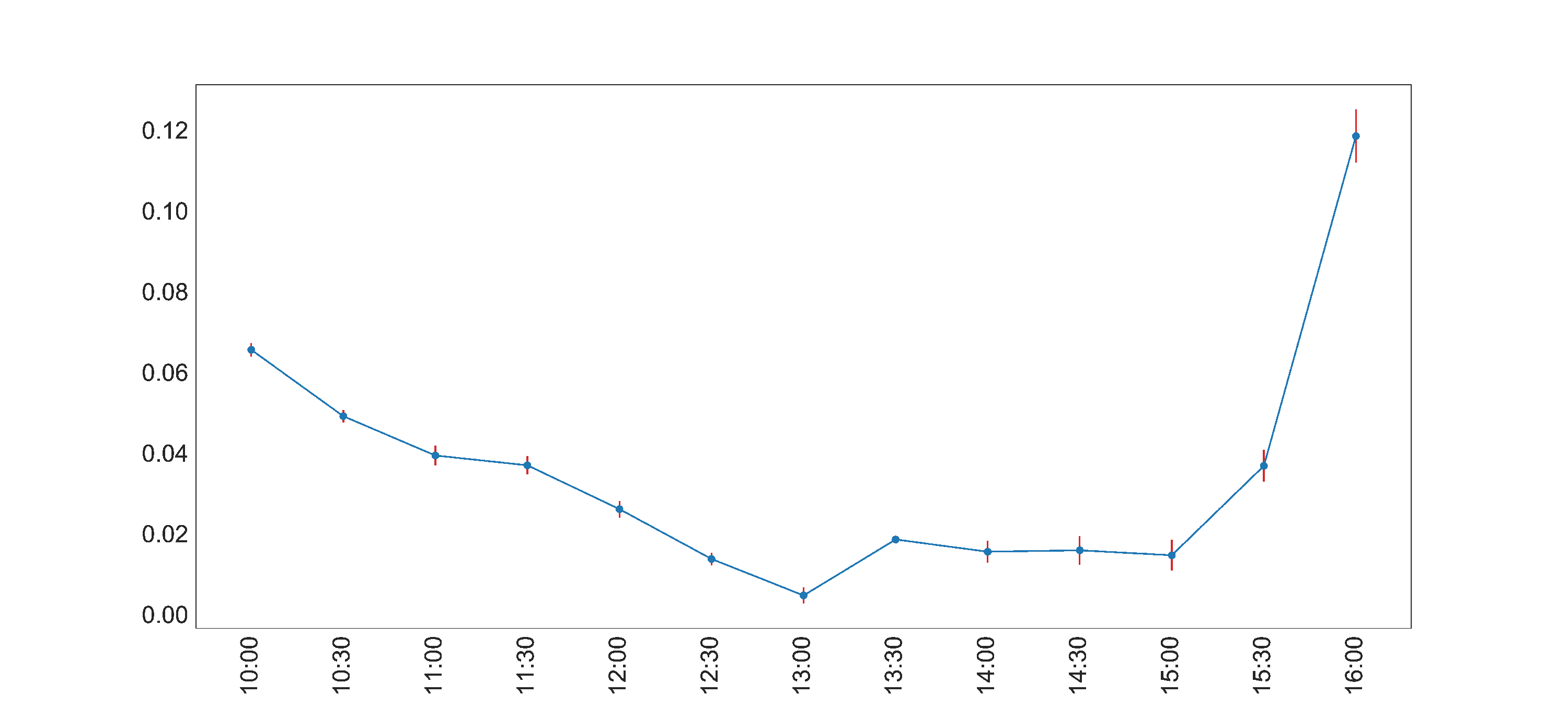}
\caption{Coefficients of the \textbf{Intraday2Daily} OLS model under {\aug}.}
\caption*{\textit{Notes:} The \textbf{Intraday2Daily} OLS model uses lagged individual 30-min RVs to forecast the next day's volatility. The $x$-axis represents the time of day. The $y$-axis represents the coefficients of lagged RVs.}
\label{fig:coef_intra2daily}
\end{figure}

For better readability, we only report the first $13=(390 / 30)$ coefficients of the OLS model using 30-min features in Figure \ref{fig:coef_intra2daily}, corresponding to the observations of RV in the most recent day.\footnote{We attain similar results for models using intraday RVs based on other frequencies.} We observe that the contributions of time-of-day dependent RVs are not even. Interestingly, \textit{volatility near the close} (15:30-16:00) is the most important predictor, in contrast to the diurnal volatility pattern. These results shed new light on the modeling of volatility.

To explain why the most recent half-hour RV is the most important predictor for forecasting the next day's volatility, we provide a handful of perspectives. According to the Wall Street Journal\footnote{\hyperlink{https://www.wsj.com/articles/the-30-minutes-that-can-make-or-break-the-trading-day-11583886131?reflink=desktopwebshare_permalink}{The 30 Minutes That Can Make or Break the Trading Day}}, there is a significant fraction of the total daily trading volume in the last half-hour of the trading day. For example, for the first few months of 2020 in the US equity market, about 23\% of trading volume in the 3,000 largest stocks by market value has taken place after 15:30. We also conclude from Figure \ref{fig:common_intra_vol} that the market achieves the highest level of consensus near the close. Therefore, volatility near the close in the previous trading day might contain more  useful information  for predicting the next day's volatility.

\section{Conclusion}\label{sec:conclusion}
In this paper, the commonality in intraday volatility over multiple horizons across the U.S. equity market is studied. By  leveraging the information content of commonality, we have  demonstrated that for most machine learning models in our analysis, pooling stock data together ({\univ}) and adding the market volatility as an additional predictor ({\aug}) generally improves the out-of-sample performance, in comparison with asset-specific models ({\single}).

We show that neural networks achieve superior performance,  possibly due to their ability to uncover and model complex  interactions among predictors. To alleviate concerns of  overfitting, we perform a stringent out-of-sample test, applying the existent trained models to unseen stocks, and conclude that  neural networks still outperform traditional models.

Lastly and perhaps most importantly, motivated by the high commonality in intraday volatility,  we propose a new approach (\textbf{Intraday2Daily}) to forecast daily RVs using past intraday RVs. The empirical findings suggest that the proposed  \textbf{Intraday2Daily} approach generally yields superior out-of-sample forecasts. We further examine the coefficients in \textbf{Intraday2Daily} OLS models, and the results suggest that volatility near the close (15:30-16:00) in the previous day (lag=1) is the most important predictor.

\paragraph{Future research directions.} 
There are a number of interesting avenues to explore in future research. One direction pertains to the assessment of whether other characteristics, such as sector RVs, can improve the forecast of future realized volatility, since in the present work, we have only considered the individual and market RVs.  Another interesting direction is to apply the underlying idea of \textbf{Intraday2Daily} approach to other risk metrics, e.g. Value-at-Risk, {that could potentially benefit from time-of-day dependent features.}


\appendix
\renewcommand\thefigure{\thesection.\arabic{figure}}    
\renewcommand\thetable{\thesection.\arabic{table}}   
\setcounter{figure}{0}    
\setcounter{table}{0}
\section{What may drive commonality in volatility?}\label{app:common_factor}

Previous studies, especially in the behavioural finance field, have shown that investor sentiments could affect stock prices (e.g. \citet{baker2007investor, kogan2006price, da2011search,bollerslev2018risk,da2015sum, karolyi2012understanding, hameed2010stock}). \citet{keynes2018general} argued that animal spirits affect consumer confidence, {thereby moving} prices in times of high levels of uncertainty. \citet{de1990noise,shleifer1990noise,kogan2006price} found that investor sentiments induce excess volatility. \citet{karolyi2012understanding} considered the investor sentiment index as an important source of commonality in liquidity. \citet{bollerslev2018risk} found a monotonic relationship between volatility and sentiment, possibly driven by correlated trading. In this section, we are interested in the relation between investor sentiments and commonality in volatility.

Traditionally, there are two approaches to measuring investor sentiments (see \citet{da2015sum}), i.e. market-based measures and survey-based indices. Following \citet{baker2007investor}, we consider the daily market volatility index (VIX) from Chicago Board Options Exchange to be the market sentiment measure. We use the Consumer Sentiment Index (CSI)\footnote{http://www.sca.isr.umich.edu} by the University of Michigan's Survey Research Center as a proxy for survey-based indices (see \citet{carroll1994does,lemmon2006consumer}). Generally speaking, CSI is a consumer confidence index, calculated by subtracting the percentage of unfavorable consumer replies from the percentage of favorable ones. Following \citet{da2015sum}, we also include a news-based index EPU\footnote{https://www.policyuncertainty.com} proposed by \citet{baker2016measuring} to measure policy-related economic uncertainty.

As suggested by \citet{morck2000information}, the raw monthly commonality measures $R^{2}_{(h), m}$ (computed based on Eqn \eqref{eqn:common_RV}) are inappropriate to use as the dependent variable in regressions, because they are bounded by 0 and 1. Consistent with \citet{morck2000information, karolyi2012understanding} and \citet{dang2015commonality}, we take the logistic transformation of $R^{2}_{(h), m}$, i.e. $\operatorname{log}\left[R^{2}_{(h), m}/(1-R^{2}_{(h), m})\right]$, denoted by $(R^{2}_{(h), m})_L$, in our following empirical analysis. To explain the commonality in volatility, we regress $(R^{2}_{(h), m})_L$ against the aforementioned three indices, as shown in Eqn \eqref{eqn:RV_ana}

\vspace{-4mm}
\begin{equation}\label{eqn:RV_ana}
(R^{2}_{(h), m})_L = \alpha + \beta_1 \text{CSI}_m + \beta_2 \text{VIX}_m + \beta_3 \text{EPU}_m +  \epsilon_{i,t}.\\
\end{equation}

Table \ref{tab:factor_common} reports the estimation results. First, we notice that a large proportion of the variance for the commonality is explained by these three sentiment factors. For example, the commonality for the 1-day scenario is 51.6\%. In terms of intraday scenarios, the R-squared values for 30-min and 65-min horizons are slightly small, 48.6\% and 48.1\%, respectively. The results {on} 10-min data are somewhat surprising, where the R-squared reaches to 55.6\%. One possible reason is that economic policy uncertainty is significant in the 10-min scenario. In another unreported robustness test, we estimate the regressions without the EPU factor. The adjusted $R^2$ value in the regression of 10-min data declines 2.5\% while for other regressions, the changes in adjusted $R^2$ are subtle. 

\begin{table}[H]
\centering
\caption{Time series regression of commonality.}
\resizebox{0.56\textwidth}{!}{\begin{tabular}{ccccc}
    \toprule
    & 10-min  & 30-min  & 65-min & 1-day   \\
    \midrule
VIX         & $0.233^*$  &$ 0.196^* $ & $0.192^*$ & $0.714^*$  \\
            & $(0.030)$   & $(0.024)$   & $(0.023)$  & $(0.084) $   \\
CSI         & $0.214^*$  & $0.097^*$  & $0.066^*$ & $0.237^*$  \\
            & $(0.025)$   & $(0.020) $  & $(0.019)$  &$ (0.070) $  \\
EPU         & $0.079^* $ & $0.025 $ & $0.022$ & $0.114$  \\
            & $(0.029)$   & $(0.023) $  & $(0.022)$  & $(0.080)$    \\ 
Constant    &$0.161^*$ & $0.982^*$ & $1.267^*$ & $-0.689^*$ \\
            & $(0.023)$   & $(0.018)$   &$ (0.018) $ &$(0.063)$   \\
            \midrule
Adjusted $R^2$ (\%) &$55.6 $& $48.6 $  & $48.1$  & $51.6 $ \\
    \bottomrule
\end{tabular}}
\label{tab:factor_common}
\caption*{\textit{Notes:} The table reports the results of time series regressions of average commonality in volatility $(R^{2}_{(h), m})_L$ over different horizons against three sentiment measures, VIX, CSI, and EPU. Superscript * denotes the significance levels of 5\%. To compare the effects of various investor sentiments, we normalize each explanatory variable by removing its mean and scaling to the unit variance.}
\end{table}

Besides the market volatility (VIX), we also find a significant effect of consumer sentiment (CSI) on the commonality of volatility over every studied horizon. The level of commonality is higher in times of higher market volatility and consumer sentiments. In addition, we observe that the coefficients of VIX and CSI for commonality in intraday volatility (especially for 30-min, 65-min) are substantially smaller than those in the daily case.

\setcounter{figure}{0}    
\setcounter{table}{0}
\section{Hyperparameter tuning} \label{app:hyperparameter}
There is no hyperparameter to tune in HAR-D and OLS. For LASSO, we use the standard 5-fold cross-validation method to determine $\lambda_1$. Hyperparameters for other models in the main analysis are summarized as follows.

\begin{table}[H]
    \centering
    \caption{Hyperparameters in XGBoost, MLP, LSTM.}
    \resizebox{0.54\textwidth}{!}{    \begin{tabular}{cccc}
    \toprule
         &XGBoost & MLP & LSTM \\
    \midrule
        Learning rate&0.1 & 0.001 & 0.001\\
        Early stopping rounds&10 & 10 & 10 \\
        Ensemble & 2000 & 10 & 10 \\
        Max depth&10 & - & - \\
        Batch size & - & 1024 & 1024 \\
        Epoches & - & 100 & 100\\
        No. of hidden layers & - & 3 & 2 \\
        Batch normalization & - & \cmark & \xmark \\
    \bottomrule
    \end{tabular}}
    \label{tab:hyperparameters}
\end{table}

{To assess the robustness of neural networks to different architectures, we repeat the main analysis using 1, 2, and 3 hidden layers.\footnote{The number of neurons is chosen based on the geometric pyramid rule, following \cite{gu2020empirical}.} The results reported in Table \ref{tab:nn_layers} are generally consistent with those reported in Table \ref{tab:experiment_sigmetaaug}.}

\begin{table}[H]
    \centering
    \caption{Out-of-sample performance of alternative hyperparameters in NNs.}
    \resizebox{0.9\textwidth}{!}{\begin{threeparttable}   
\begin{tabular}{l l c c c c c c c c }
    \toprule
  &  & \multicolumn{2}{c}{{10-min}} & \multicolumn{2}{c}{{30-min}} & \multicolumn{2}{c}{{65-min}} & \multicolumn{2}{c}{{1-day}} \\
     \cmidrule(lr){3-4}\cmidrule(lr){5-6}\cmidrule(lr){7-8}\cmidrule(lr){9-10}
&  & MSE & QLIKE & MSE & QLIKE & MSE & QLIKE & MSE & QLIKE \\ \midrule
\multirow{2}{*}{MLP1} & {\univ}  & 0.949 & 0.398 & 0.286 & 0.182 & 0.234 & 0.164 & 0.261 & 0.191 \\
                      & {\aug}   & 0.947 & 0.388 & 0.281 & 0.181 & 0.229 & 0.162 & 0.257 & 0.181 \\\midrule
\multirow{2}{*}{MLP2} & {\univ}  & 0.948 & 0.398 & 0.284 & 0.182 & 0.232 & 0.164 & 0.260 & 0.190  \\
                      & {\aug}   & 0.947 & 0.387 & 0.281 & 0.180 & 0.229 & 0.163 & 0.256 & 0.180 \\\midrule
\multirow{2}{*}{MLP3} & {\univ}  & 0.947 & 0.397 & 0.284 & 0.181 & 0.232 & 0.163 & 0.260 & 0.191 \\
                      & {\aug}   &{0.945} &{0.386} &{0.280} &{0.179} &{0.229} &{0.162} & 0.257 & 0.180 \\\midrule
\multirow{2}{*}{LSTM1} & {\univ} & 0.956 & 0.398 & 0.293 & 0.190 & 0.232 & 0.163 & 0.262 & 0.189 \\
                       & {\aug} & 0.938 & 0.383 & {0.286} &{0.181} &{0.230} &{0.161} & 0.259 & 0.182 \\\midrule
\multirow{2}{*}{LSTM2} & {\univ} & 0.950 & 0.393 & 0.287 & 0.179 & 0.232 & 0.162 & 0.261 & 0.188 \\
                       & {\aug} &{0.934} &{0.376} &{0.279} &{0.171} &{0.229} &{0.160} & 0.258 & 0.182\\\midrule
\multirow{2}{*}{LSTM3} & {\univ}  & 0.949 & 0.392 & 0.286 & 0.178 & 0.232 & 0.163 & 0.260 & 0.187 \\
                       & {\aug} & 0.933 & 0.376 & 0.280 & 0.171 & 0.229 & 0.161 & 0.256 & 0.181 \\
 
\bottomrule
\end{tabular}

\end{threeparttable}   }
    \caption*{\textit{Notes:} MLP1 has single hidden layer with 128 neurons. MLP2 has two hidden layers of 128 and 64 neurons, respectively. MLP3 has three hidden layers of 128, 64 and 32 neurons, respectively. LSTM variants have similar meanings.}
    \label{tab:nn_layers}
\end{table}

\setcounter{figure}{0}    
\setcounter{table}{0}
\section{Diebold-Mariano test}\label{app:DM}
Diebold-Mariano (DM) test is used to discriminate the significant differences of forecasting accuracy between different time series models (for example \citet{diebold1995comparing, diebold2015comparing}). Denote the loss associated with forecast error $e_{t}$ by $L(e_{t})$, e.g. $L(e_{t})=e_{t}^{2}$. Then the loss difference between the forecasts of models $a$ and $b$ is given by  $d_{t}^{(a-b)}=L(e_{t}^{(a)})-L(e_{t}^{(b)}),$ where $e_{t}^{(a)}$ ($e_{t}^{(b)}$) represents the forecast error from model $a$ ($b$), respectively. The DM test makes one assumption that $d_{t}^{(a-b)}$ is covariance stationary. The null hypothesis is that $\mathbb{E}(d_{t}^{(a-b)})=0$. Under the covariance stationary assumption, we have the test statistic
\vspace{-3mm}
\begin{equation}
DM_{12}=\frac{\bar{d}^{(a-b)}}{\hat{\sigma}^{(a-b)}} \rightarrow N(0,1),
\end{equation}
where $\bar{d}^{(a-b)}=\frac{1}{T} \sum_{t=1}^{T} d_{t}^{(a-b)}$ is the sample mean of $d_{t}^{(a-b)}$,  and $\hat{\sigma}^{(a-b)}$ is a consistent estimate of the standard deviation of $\bar{d}^{(a-b)}$.

Following \citet{gu2020empirical}, we apply a modified {DM test}, to make pairwise comparisons of models' performance when forecasting multi-asset volatility. Specifically, the modified DM test compares the cross-sectional average of prediction errors from each model, rather than comparing errors for each individual asset, i.e.
\begin{equation}
d_{t}^{(a-b)}=\frac{1}{N} \sum_{i=1}^N \left(L(e_{i, t}^{(a)})-L(e_{i, t}^{(b)})\right),
\end{equation}
where $e_{i, t}^{(a)}$ ($e_{i, t}^{(b)}$)  refers to the forecast error for stock $i$ at time $t$ from model $a$ ($b$), respectively.

To assess the statistical significance of the differences in out-of-sample volatility forecasts as shown in Table \ref{tab:experiment_sigmetaaug}, we report the results of all DM tests in terms of QLIKE for each horizon.

\begin{table}[H]
	\caption{Statistics of Diebold-Mariano tests.}
		\label{tab:DM}
\end{table}

\vspace{-8mm}
\begin{table}[H]
\centering
\caption*{Panel A: 10-min.}
    \resizebox{1\textwidth}{!}{\begin{tabular}{c|cccccc}
   \hline
\diagbox[width=5em]{Univ}{Univ} & LASSO   & OLS  & LASSO & XGBoost & MLP & LSTM \\
   \hline
LASSO   &&$ 42.33^* $&$ 36.17^*  $&$ 56.55^* $&$ 83.26^* $&$ 72.43^*$\\
OLS     &&&$ -32.84^* $&$ 33.30^* $&$ 62.29^* $&$ 52.90^*$ \\
Lasso   &&&&$ 35.00^* $&$ 62.15^* $&$ 54.17^* $\\
XGBoost &&&&&$ 25.86^* $&$ 20.31^* $\\
MLP     &&&&&&$ -3.39^* $\\
LSTM    &&&&&&  \\
   \hline
Single vs  &$-30.07^*$   &$-1.02$    &  $31.27^*$   &$59.38^*$     &     &      \\
   \hline
\end{tabular}
\begin{tabular}{c|cccccc}
   \hline
\diagbox[width=5em]{Aug}{Aug} & LASSO   & OLS  & LASSO & XGBoost & MLP & LSTM \\
   \hline
LASSO   &&$ 56.60^* $&$ 58.95^* $&$ 33.75^* $&$ 68.81^* $&$ 66.94^* $\\
OLS     &&&$ 5.69^*  $&$ -6.02^* $&$ 31.82^* $&$ 31.71^* $\\
Lasso   &&&&$ -6.52^* $&$ 30.99^* $&$ 31.43^* $\\
XGBoost &&&&&$ 21.54^* $&$ 28.72^* $\\
MLP     &&&&&&$ 14.51^* $\\
LSTM    &&&&&& \\
   \hline
Univ vs  &$46.23^*$   &$51.28^*$    &$53.53^*$     &$-0.32$     &$6.24^*$     &$29.63^*$      \\
   \hline
\end{tabular}}
\label{tab:DM_metaaug10min}
\end{table}

\vspace{-3mm}
\begin{table}[H]
\centering
\caption*{Panel B: 30-min.}
    \resizebox{1\textwidth}{!}{\begin{tabular}{c|cccccc}
   \hline
\diagbox[width=5em]{Univ}{Univ} & HAR-D   & OLS  & LASSO & XGBoost & MLP & LSTM \\
   \hline
HAR-D   &   & $44.74^*$   & $44.00^* $   & $42.96^*$    & $52.99^*  $ & $46.17^* $  \\
OLS     &  &   & $-23.57^*$   & $22.63^*$    & $35.25^*$   & $26.19^*  $ \\
LASSO   & &  &        & $24.55^*$    & $37.07^* $  & $28.04^* $  \\
XGBoost &  & &   &     &$ 16.46^* $  & $8.35^*$    \\
MLP     &   & &  &  &     & $-8.72^* $  \\
LSTM    & &  &  & & &  \\
   \hline
Single vs  &$-7.47^*$   &$-0.48$    &   $23.14^*$  &$48.57^*$     &    & \\
   \hline
\end{tabular}

\begin{tabular}{c|cccccc}
   \hline
\diagbox[width=5em]{Aug}{Aug} & HAR-D   & OLS  & LASSO & XGBoost & MLP & LSTM \\
   \hline
HAR-D   &   & $36.36^*$    & $38.00^* $   & $24.16^* $   & $45.12^*$   & $44.43^* $  \\
OLS     & &   & $-2.11^* $ & $-4.50^*  $  & $20.55^*$   & $18.26^*$   \\
LASSO   & &   &    & $-4.34^* $   & $21.21^*$   & $19.05^*$   \\
XGBoost & &  &  &   & $21.04^*$   &$ 23.02^*$   \\
MLP     & & &  & &  & $2.24^* $ \\
LSTM    &  & & &  & & \\
   \hline
Univ vs  &$17.70^*$   &$22.51^*$    &$25.24^*$     &$-9.59^*$     &$9.56^*$     &$23.23^*$      \\
   \hline
\end{tabular}}
\label{tab:DM_metaaug30min}
\end{table}

\vspace{-3mm}
\begin{table}[H]
\caption*{Panel C: 65-min.}
\centering
    \resizebox{1\textwidth}{!}{\begin{tabular}{c|cccccc}
   \hline
\diagbox[width=5em]{Univ}{Univ} & HAR-D   & OLS  & LASSO & XGBoost & MLP & LSTM \\
   \hline
HAR-D   & & $28.27^*$  & $27.75^*$   & $21.56^*$  & $30.06^*$  & $29.00^*$  \\
OLS     & &     & $-11.73^*$  & $7.78^*$   & $20.22^*$  & $18.67^*$  \\
LASSO   & &     &      & $8.81^*$   & $20.91^*$  & $19.35^*$  \\
XGBoost & &     &      &     & $19.83^*$  & $18.17^*$  \\
MLP     & &     &      &     &     & $0.68^*$   \\
LSTM    & &     &      &     &     &     \\
   \hline
Single vs  &$-1.87$   &$8.17^*$    &    $8.53^*$  &$41.26^*$     &    &    \\
   \hline
\end{tabular}
\begin{tabular}{c|cccccc}
   \hline
\diagbox[width=5em]{Aug}{Aug} & HAR-D   & OLS  & LASSO & XGBoost & MLP & LSTM \\
   \hline
HAR-D   & & $22.11^*$ & $22.67^*$ & $9.87^*$  & $25.92^*$ & $26.01^*$ \\
OLS     & &     & $-4.89^*$ & $-5.56^*$ & $12.84^*$ & $11.79^*$ \\
LASSO   & &     &     & $-5.11^*$ & $13.72^*$ & $12.67^*$ \\
XGBoost & &     &     &     & $17.71^*$ & $18.54^*$ \\
MLP     & &     &     &     &     & $0.94^*$  \\
LSTM    & &     &     &     &     &    \\
   \hline
Univ vs  &$10.92^*$   &$12.47^*$    &$13.35^*$     &$-7.52^*$     &$7.15^*$     &$7.94^*$      \\
   \hline
\end{tabular}}
\label{tab:DM_metaaug65min}
\end{table}

\vspace{-3mm}
\begin{table}[H]
\caption*{Panel D: 1-day.}
\centering
    \resizebox{1\textwidth}{!}{\begin{tabular}{c|cccccc}
   \hline
\diagbox[width=5em]{Univ}{Univ} & HAR-D   & OLS  & LASSO & XGBoost & MLP & LSTM \\
   \hline
HAR-D   & &$ 0.50 $&$ -0.12 $&$ -4.29^* $&$ 2.69^* $&$ 3.86^*  $\\
OLS     & & &$ -4.94^* $&$ -5.50^* $&$ 1.91$&$ 3.36^*  $\\
LASSO   & & & &$ -4.29^* $&$ 2.76^* $&$ 3.88^*  $\\
XGBoost & & & & &$ 9.49^* $&$ 10.90^* $\\
MLP     & & & & & &$ 3.03^*  $\\
LSTM    & & & & & &   \\
   \hline
Single vs  &$-1.32$   &$3.41^*$    &   $5.42^*$ &$20.53^*$     &     &     \\
   \hline
\end{tabular}
\begin{tabular}{c|cccccc}
   \hline
\diagbox[width=5em]{Aug}{Aug} & HAR-D   & OLS  & LASSO & XGBoost & MLP & LSTM \\
   \hline
HAR-D   & &$ -0.12 $&$ 5.39^*  $&$ -8.97^*  $&$ 3.20^*  $&$ 1.87  $\\
OLS     & & &$ -1.10 $&$ -12.63^* $&$ -3.77^* $&$ -3.99^* $\\
LASSO   & & &$       $&$ -12.68^* $&$ -3.34^* $&$ -3.91^* $\\
XGBoost & & & & &$ 12.30^* $&$ 11.61^* $\\
MLP     & & & & & &$ -2.20* $\\
LSTM    & & & & & &     \\  
   \hline
Univ vs  &$4.50^*$   &$5.81^*$    &$6.30^*$     &$-5.01^*$     &$4.63^*$     &$2.78^*$      \\
   \hline
\end{tabular}}
\label{tab:DM_metaaug1day}

\caption*{\textit{Notes:} In each panel, the left sub-table represents the pairwise comparison of forecasting performance of six models trained under {\univ} and the right one represents the pairwise comparison of forecasting performance of six models trained under {\aug}. The bottom row in each sub-table represents the comparison of forecasting performance of the same model under two different training schemes. Positive numbers indicate the column model outperforms the row model. Superscript * denotes the significance levels of 5\%.}
\end{table}

\setcounter{figure}{0}    
\setcounter{table}{0}
\section{Model update frequency} \label{app:model_update}
In the present paper, we choose to update each risk model annually due to the limited computation resources. To understand whether the model's performance might change with respect to the update frequency, we update the HAR-D model with different frequencies, i.e. weekly, monthly, and yearly, and results are summarized as follows. The conclusions are generally consistent with those from our main analysis.
\begin{table}[H]
	\caption{Frequency of updating HAR-D for predicting intraday RVs.}
	\centering
    \resizebox{0.95\textwidth}{!}{\begin{tabular}{l l c c c c c c c c }
    \toprule
  \multicolumn{2}{c}{\textbf{ Panel A:} }  & \multicolumn{2}{c}{{10-min}} & \multicolumn{2}{c}{{30-min}} & \multicolumn{2}{c}{{65-min}} & \multicolumn{2}{c}{{1-day}} \\
     \cmidrule(lr){3-4}\cmidrule(lr){5-6}\cmidrule(lr){7-8}\cmidrule(lr){9-10}
     
 \multicolumn{2}{c}{\textbf{ Statistical performance} } & MSE & QLIKE & MSE & QLIKE & MSE & QLIKE & MSE & QLIKE \\\hline

\multirow{3}{*}{Weekly} &{\single}   &1.013 &0.483 &0.332 &0.221 &0.270 &0.190 &0.267 &0.188  \\
                        &{\univ}  &1.021 &0.517 &0.333 &0.230 &0.270 &0.190 &0.268 &0.189\\
                        &{\aug}  &0.995 &0.453 &0.323 &0.228 &0.262 &0.185 &0.256 &0.180  \\ \hline
\multirow{3}{*}{Monthly} &{\single}   &1.013 &0.483 &0.332 &0.222 &0.270 &0.190 &0.267 &0.189  \\
                        &{\univ}  &1.021 &0.517 &0.333 &0.230 &0.270 &0.191 &0.268 &0.190 \\
                        &{\aug}  &0.995 &0.453 &0.323 &0.227 &0.262 &0.185 &0.256 &0.180  \\ \hline
\multirow{3}{*}{Yearly} &{\single}   &1.013 &0.484 &0.332 &0.222 &0.270 &0.190 &0.269 &0.190  \\
                        &{\univ}  &1.021 &0.518 &0.333 &0.230 &0.270 &0.191 &0.269 &0.190 \\
                        &{\aug}  &0.995 &0.453 &0.323 &0.227 &0.262 &0.186 &0.257 &0.180  \\ \midrule
  \multicolumn{2}{c}{\textbf{ Panel B:} }  & \multicolumn{2}{c}{{10-min}} & \multicolumn{2}{c}{{30-min}} & \multicolumn{2}{c}{{65-min}} & \multicolumn{2}{c}{{1-day}} \\
     \cmidrule(lr){3-4}\cmidrule(lr){5-6}\cmidrule(lr){7-8}\cmidrule(lr){9-10}
     
 \multicolumn{2}{c}{\textbf{ Realized utility} } & RU & RU-TC & RU & RU-TC & RU & RU-TC & RU & RU-TC \\\hline

\multirow{3}{*}{Weekly} &{\single}   &2.694 &2.069 &3.459 &3.042 &3.543 &3.096 &3.551 &3.518  \\
                        &{\univ}  &2.575 &1.972 &3.427 &3.014 &3.541 &3.095 &3.548 &3.516\\
                        &{\aug}  &2.790 &2.280 &3.427 &3.020 &3.553 &3.108 &3.571 &3.536  \\ \hline
\multirow{3}{*}{Monthly} &{\single}   &2.693 &2.068 &3.458 &3.042 &3.542 &3.096 &3.549 &3.516  \\
                        &{\univ}  &2.574 &1.972 &3.427 &3.014 &3.541 &3.095 &3.547 &3.514\\
                        &{\aug}  &2.789 &2.279 &3.426 &3.020 &3.553 &3.107 &3.571 &3.536  \\ \hline
\multirow{3}{*}{Yearly} &{\single}   &2.690 &2.065 &3.457 &3.040 &3.542 &3.095 &3.548 &3.515  \\
                        &{\univ}  &2.574 &1.975 &3.429 &3.016 &3.541 &3.095 &3.547 &3.514\\
                        &{\aug}  &2.790 &2.280 &3.428 &3.022 &3.552 &3.107 &3.571 &3.536  \\ 
    \bottomrule
    \end{tabular}
}
	\label{tab:model_update}
\end{table}

\end{document}